\documentclass[final,5p,times]{elsarticle}
\usepackage{graphicx}
\usepackage{amssymb}
\usepackage{pdfpages}

\usepackage[colorlinks=true,citecolor=blue,linkcolor=blue,urlcolor=blue]{hyperref}
\usepackage[inline]{enumitem}
\usepackage{bm}
\usepackage{amsmath} % \text, and other math formatting options
\usepackage{mathtools}
\usepackage[normalem]{ulem}

\graphicspath{{figures/}{./}}

\newcommand{\fref}[1]{Fig.~\ref{#1}}
\newcommand{\tref}[1]{Table~\ref{#1}}
\newcommand{\eref}[1]{Eq.~\eqref{#1}}
\newcommand{\sref}[1]{Section~\ref{#1}}
\newcommand{\act}[0]{h}
\newcommand{\correction}[1]{{\color{black}  #1}}

\biboptions{sort&compress}

\newcounter{bla}

\makeatletter
\def\ps@pprintTitle{%
  \let\@oddhead\@empty
  \let\@evenhead\@empty
  \let\@oddfoot\@empty
  \let\@evenfoot\@oddfoot
}
\makeatother

\begin{document}

\begin{frontmatter}



\title{KLIFF: A framework to develop physics-based and machine learning interatomic potentials}


\author[a]{Mingjian Wen\fnref{author}}
\author[a]{Yaser Afshar}
\author[a]{Ryan S.\ Elliott}
\author[a]{Ellad B.\ Tadmor\corref{author}}

\fntext[author]{Current address:  Energy Technologies Area, Lawrence Berkeley National Laboratory, Berkeley, CA 94720, United States.}
\cortext[author] {Corresponding author. \textit{E-mail address:} tadmor@umn.edu}
\address[a]{Department of Aerospace Engineering and Mechanics, University of Minnesota, Minneapolis, MN 55455, USA}

\begin{abstract}
%A submitted program is expected to be of benefit to other physicists or physical chemists, or be an exemplar of good programming practice, or illustrate new or novel programming techniques which are of importance to some branch of computational physics or physical chemistry.

%Acceptable program descriptions can take different forms. The following Long Write-Up structure is a suggested structure but it is not obligatory. Actual structure will depend on the length of the program, the extent to which the algorithms or software have already been described in literature, and the detail provided in the user manual.

%Your manuscript and figure sources should be submitted through the Elsevier Editorial System (EES) by using the online submission tool at \\
%http://www.ees.elsevier.com/cpc.

%In addition to the manuscript you must supply: the program source code; job control scripts, where applicable; a README file giving the names and a brief description of all the files that make up the package and clear instructions on the installation and execution of the program; sample input and output data for at least one comprehensive test run; and, where appropriate, a user manual. These should be sent, via email as a compressed archive file, to the CPC Program Librarian at cpc@qub.ac.uk.

Interatomic potentials (IPs) are reduced-order models for calculating the potential
energy of a system of atoms given their positions in space and species.
IPs treat atoms as classical particles without explicitly modeling electrons
and thus are computationally far less expensive than first-principles methods,
enabling molecular simulations of significantly larger systems over longer times.
Developing an IP is a complex iterative process involving
multiple steps: assembling a training set, designing a functional form,
optimizing the function parameters, testing model quality,
and deployment to molecular simulation packages.
This paper introduces the \emph{KIM-based learning-integrated fitting framework}
(KLIFF), a package that facilitates the entire IP development process.
KLIFF supports both physics-based and machine learning IPs. It adopts a modular approach
whereby various components in the fitting process, such as atomic environment descriptors, functional forms, loss functions, optimizers, quality analyzers, and so on, work seamlessly with
each other. This provides a flexible framework for the rapid design of new IP forms.
Trained IPs are compatible with the Knowledgebase of Interatomic Models (KIM)
application programming interface (API) and can be readily used in major
materials simulation packages compatible with KIM, including ASE,
DL\_POLY, GULP, LAMMPS, and QC\@.
KLIFF is written in Python with computationally intensive components implemented in C++.
It is parallelized over data and supports both
shared-memory multicore desktop machines and high-performance distributed memory computing clusters.
We demonstrate the use of KLIFF by fitting a physics-based Stillinger--Weber
potential and a machine learning neural network potential for silicon.
The KLIFF package, together with its documentation, is publicly available at:
\url{https://github.com/openkim/kliff}.
\end{abstract}


\end{frontmatter}

\section{Introduction}
\label{sec:intro}

%%%%%%%%%%%%%%%%%%%%
% why and what is IP
%%%%%%%%%%%%%%%%%%%%
Molecular simulations are a powerful computational technique for exploring
material behavior and properties based on an understanding of the physics of bonding
at the atomic scale \cite{tadmor2011modeling}.
This approach is used across the sciences with examples such as
phase transition in crystals \cite{khaliullin2011nucleation},
protein folding \cite{piana2012protein}, and
thermal expansion and conductivity of layered 2D materials \cite{wen2017sw, wen2019hybrid}
to name just a few.
At the core of any molecular simulation lies a description of the interactions
between atoms that produces the forces governing atomic motion.
First-principles approaches (e.g.\ density functional theory (DFT)) that involve
solving the Schr\"{o}dinger equation of quantum mechanics are most accurate,
but due to hardware and
algorithmic limitations, these approaches are limited to extremely small system
sizes and time scales precluding the study of most systems of technological interest.
%This excludes the possibility to study many problems of interest that requires
%larger length scale and/or time scale.
For example, the supercell required to simulate a graphene bilayer with a
$1.1^\circ$ twist angle has more than 10,000 atoms, which is well beyond the capabilities
of current first-principles approaches \cite{wen2019hybrid}.

Interatomic potentials (IPs, also known as force fields) provide a classical
alternative based on the Born--Oppenheimer approximation (BOA)
\cite{born1927quantentheorie}. Due to the large mass difference between nuclei and
electrons, the BOA assumes that electrons instantaneously adapt to changes in nuclei
positions adopting their ground state configuration --- effectively decoupling
nuclei and electron physics. This approximation is reasonable for many problems of
interest in materials science and condensed-matter physics \cite{tadmor2011modeling}.
Consistent with the BOA, IPs treat atoms as classical particles without explicitly
modeling the electrons, but strive to capture their influence on atomic nuclei in an
effective manner.  As such, IPs are computationally far less expensive than first-principles
methods and can therefore be used to compute static and dynamic properties that
are inaccessible to first-principles calculations \cite{mishin1999interatomic,
wen2015interpolation, wen2017potfit}.
In essence, an IP is a reduced-order model for the quantum-mechanical
interaction of electrons and nuclei in a material through a parameterized
functional form that depends only on the positions of the atomic nuclei (atoms hereafter).

\begin{figure}[!t]
\centering
\includegraphics[width=0.9\columnwidth]{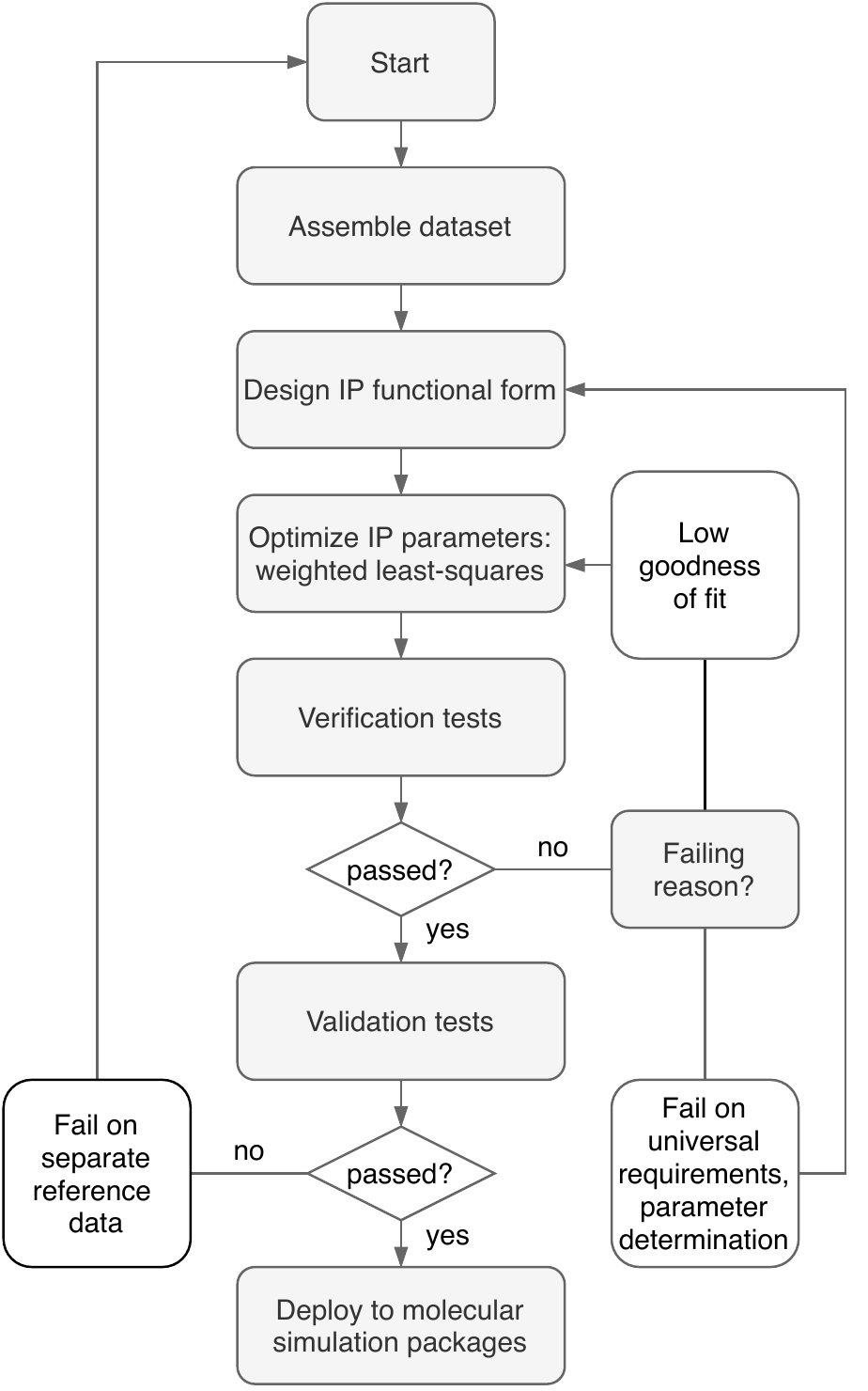}
\caption{Flowchart of the IP development process.
\correction{
Developing an IP involves four major steps:
(1) assemble a set of reference data and design an IP functional form;
(2) optimize IP parameters, typically carried out by minimizing a weighted least-squares loss function of the model predictions and the reference data;
(3) assess the quality of the optimized model via verification and validation tests;
and
(4) deploy the model to molecular simulation packages.
These steps can be iterative.
When a model fails a verification test (e.g.\ by not satisfying a universal requirement, such as translational and rotational invariance, or by having a low goodness of fit on a test set) or fails a validation test (e.g.\ being unable to reproduce experimental material properties), it is necessary to return to earlier steps, make adjustments, and redo the fitting.
}
}
\label{fig:ipdev:flowchart}
\end{figure}

%%% functional form
Development of an IP is a complex iterative process involving multiple steps
as shown in \fref{fig:ipdev:flowchart}. (Refer back to this figure as your read
the remainder of this section.)
First, a dataset of experimental and/or first principles reference data must be
assembled to which the IP will be fitted.
When developing machine learning potentials, it is common practice to split the
dataset into three parts:
(1) a \emph{training} set that is used to optimize the model parameters,
(2) a \emph{validation} set for fitting hyperparameters and monitor overfitting, and
(3) a \emph{test} set to assess the goodness of the fit.

Traditionally, the reference dataset  contains material properties considered
important for a given application, such as the cohesive energy, equilibrium
lattice constant, and elastic moduli of given crystal phases to name a few.
In recent years, many IPs adopt a \emph{force-matching} scheme
\cite{ercolessi1994interatomic}, in which the training set is augmented
with the forces on atoms obtained by first-principles calculations for a large
set of atomic configurations.\footnote{These can be configurations associated
with important structures or snapshots of the crystal as the atoms
oscillate at finite temperature or through random perturbations.}
An advantage of this approach is that the issue of insufficient training data
(particularly true for machine learning potentials) can be resolved because
as many training data as needed can be readily generated.

Construction of a good reference dataset is critical for success.
The fidelity of the IP for a given application hinges on including
the appropriate physics in the dataset. It is also important to not
swamp out rare configurations (such as transition states) that can
have a disproportionate effect on material behavior. Dataset
curation remains a difficult open problem and an area
of active research \cite{zhang2015train}.

Next an appropriate functional form has to be selected.
Traditionally, the functional form of an IP was devised to represent the physics
underlying the material system.
One of the earliest examples is the pair potential developed by
Lennard-Jones (LJ) in the 1920s to model van der Waals interactions in noble gases
\cite{jones1924a,jones1924b,lennardjones1931}. The LJ potential includes an $r^{-6}$ term
(where $r$ is the distance between atoms) that is based on a theoretical model
for London dispersion, and an $r^{-12}$ term meant to model repulsion due
to Pauli exclusion.
\correction{
In the past century, a large number of physics-based potentials have been developed
for a variety of ionic, metallic, and covalent systems \cite{tadmor2011modeling}.
A physics-based potential typically adopts a closed-form functional expression that is based
on known physical or geometric aspects of bonding in the material.}
The functional forms of these IPs have become increasingly complex with
an ever growing number of parameters.\footnote{For example, there are only two parameters in the
LJ potential \cite{lennardjones1931}, whereas the ReaxFF
\cite{duin2001reaxff} model developed for more complex systems
has hundreds of adjustable parameters.}

Devising the appropriate functional form to correctly capture the physics
underlying the material system is arguably the most difficult task in developing
a physics-based potential. It involves a mix of art and science as pointed out by
Brenner \cite{brenner2000art}.
This is largely alleviated by machine learning potentials \cite{behler2007generalized,
bartok2010gaussian, rupp2012fast, thompson2015spectral, shapeev2016moment},
which have emerged in recent years and been shown to be highly effective for a
spectrum of material systems ranging from organic molecules \cite{rupp2012fast}
to alloys \cite{hajinazar2017stratified}.
Different from physics-based potentials, machine learning potentials are constructed by first
transforming the atomic environment information in a large training set of
first-principles results into vector representations (descriptors) and then
training general-purpose regression algorithms on the atomic environment descriptors.
In a machine learning potential, the regression algorithm contains no physics,
but instead it attempt to ``learn'' the quantum mechanical
Schr\"{o}dinger equation directly from the training set of reference data.
Properly tuned with a sufficiently dense training set, machine learning potentials have the advantage that,
in principle, they can describe arbitrary bonding states and thus can achieve extremely high accuracy.

%%% fitting
After the functional form has been selected (either physics-based or machine learning), the
next step is to determine the values of the function parameters.
This is typically formulated as a least-squares minimization problem by first constructing
a \emph{loss function} that quantifies the difference between the IP predictions and the
reference values in the training set and then adjusting the parameters to reduce the loss
function as much as possible. This can be challenging because
IPs are nonlinear functions that are often ``sloppy'' in the sense that their
predictions are insensitive to certain parameters or certain combinations of
parameters \cite{waterfall:casey:2006, kurniawan2021bayesian}.
These soft modes in parameter space can cause the minimization algorithms to
fail to converge \cite{wen2017potfit}.
A solution is to use a minimization algorithm that moves along
flat regions in parameter space more quickly
(e.g.\ the geodesic Levenberg--Marquardt algorithm \cite{transtrum2011geometry,transtrum2012geodesic,transtrum2012improvements}),
or better yet, to identify soft modes using a sensitivity analysis (e.g.\ a Fisher information
based method \cite{wen2017sw}) and then apply a suitable model reduction.

%%% assessment (refinement)
Once an IP is trained, its quality must be assessed.
This can be approached from a verification \& validation (V\&V) perspective.
These terms are defined as \cite{TMSVV2019}:
\begin{itemize}
\item \textbf{Verification:} The process of determining that a computational model accurately represents the underlying mathematical model and its solution.
\item \textbf{Validation:}\footnote{Note that the term \emph{validation} is used differently in the V\&V context than the validation set in machine learning mentioned above.} The process of determining the degree to which a model is an accurate representation of the real world from the perspective of the intended uses of the model.
\end{itemize}
Verification for an IP includes satisfaction of universal requirements such as
translational and rotational invariance (objectivity), permutation symmetry,
forces returned by the IP correspond to the negative gradient of the energy,
and so on. These are referred to as ``verification checks'' within the
Knowledgebase of Interatomic Models (KIM) framework
\cite{tadmor2011kim,tadmor2013nsf,karls:bierbaum:2020,openkim}.
In addition, verification includes tests that assess the
quality of the model in terms of the uncertainty in parameter determination,
and the goodness of the fit using a test set as mentioned above.

The V\&V notion of validation can be understood within the context of
\emph{transferability}, i.e.\ the ability of the IP to predict phenomena that it
was not fit to reproduce. This includes prediction of material properties,
computed by ``KIM Tests'' within the KIM framework \cite{karls:bierbaum:2020},
and predictions obtained through large-scale molecular simulations of real-world behavior.
For example, the ability of IPs for carbon
to reproduce the experimental structure of amorphous carbon \cite{detomas2016acarbon}.

As a general rule, physics-based potentials are better placed to exhibit transferability than machine
learning potentials as long as the functional forms capture the requisite physics. For example an
LJ potential fitted to the properties of an ideal gas provides a good approximation
(within 10\%) for the ground state crystal structure obtained by cooling the gas down to 0~K
\cite{tadmor2011modeling}. This is an impressive demonstration of transferability. In contrast,
machine learning potentials have no physics beyond that in the training set (and possibly the
descriptors). This means that a machine learning potential can only ``transfer'' to configurations
that are close to what already exists in its training set.

Transferability can be included in the IP fitting process through a
comparison of IP predictions with separate reference data.
In cases where this fails, either the functional form needs to be extended for
a physics-based potential, and/or the training set needs to be expanded for both
physics-based and machine learning potentials. The training must then be redone.

%%%% deployment
Finally, once the IP fitting process is complete, the IP must be deployed to one or more
molecular simulation packages of choice. Traditionally this is done on a code-by-code basis,
which can be a time consuming and error prone process.
If the IP class is already implemented in the code, then simply providing parameters may
be enough --- although even there things can go wrong. For example, for the REBO potential
\cite{brenner2002second} implemented in LAMMPS \cite{lammps}, some of
the parameters were not the ones presented in the original paper by Brenner
\emph{et al.} \cite{brenner2002second}, but rather from the closely related
AIREBO potential \cite{stuart2000airebo}.\footnote{This has been corrected in more recent
implementations.}
In situations where an IP class is not available in a simulation code, the work involved in
implementing it may be prohibitive. For example in the amorphous carbon study mentioned above
\cite{detomas2016acarbon} only IPs implemented in LAMMPS were tested, leaving out more than
half of the possible IPs identified by the authors.
The KIM application programming interface (API) \cite{kimapi} was designed to address this
by creating a standard that allows a conforming IP to work seamlessly with any simulation
code that supports it. The KIM API is supported by major materials simulation platforms
including ASE \cite{larsen2017atomic,ase}, DL\_POLY \cite{smith1996,dlpoly},
GULP \cite{gale1997gulp,gulp}, LAMMPS \cite{plimpton1995fast,lammps2021,lammps}, and QC \cite{tadmor1996,qc}.

%Recent efforts such as the NIST interatomic potentials repository \cite{nist},
%the Atomistica interatomic potentials repository \cite{atomistica}, and the
%Open Knowledgebase of Interatomic Models (KIM) project
%\cite{tadmor2011kim,tadmor2013nsf,openkim} strive to archive parameter files
%as well as the implementations of a variety of IPs.
%IP implementations in KIM are stored in a format that conforms to an application
%programming interface (API) standard developed as part of the KIM project
%\cite{kimapi}.
%The KIM API enables any IP stored in KIM to work seamlessly with any
%KIM-compliant simulation package including LAMMPS \cite{plimpton1995fast,lammps},
%ASE \cite{larsen2017atomic,ase}, DL\_POLY \cite{smith1996,dlpoly}, and GULP \cite{gale1997gulp,gulp} among others.
%

This paper introduces the KIM-based learning-integrated fitting framework (KLIFF),
a package that facilitates the entire IP development process described above.
KLIFF provides a unified Python interface to train both physics-based and machine
learning potentials, and is constructed in modular fashion, making it easy to use
and extend.
It integrates closely with the KIM ecosystem for accessing
IPs to train, testing trained IPs, and deploying trained IPs.
The paper is structured as follows.
\sref{sec:ip} introduces two example IPs (one physics-based and the other machine learning)
that will be trained later,
and discusses the least-squares approach used to parameterize IPs.
\sref{sec:features} presents KLIFF features and capabilities.
Implementation details of the code are outlined in \sref{sec:implementation}.
\sref{sec:demo} presents a demonstration of using KLIFF to fit the two IPs introduced in
\sref{sec:ip}. The paper concludes in \sref{sec:summary} with a summary.

%%%%%%%%%%%%%%%%%%%%%%%%%%%%%%%%%%%%%%%%%%%%%%%%%%%%%%%%%%%%%%%%%%%%%%%%%%%%%%%%
% IP
%%%%%%%%%%%%%%%%%%%%%%%%%%%%%%%%%%%%%%%%%%%%%%%%%%%%%%%%%%%%%%%%%%%%%%%%%%%%%%%%
\section{Interatomic potentials}
\label{sec:ip}

An IP is typically formulated as a parametric model that takes the positions of
the atoms as its arguments and returns the potential energy,\footnote{In general, IPs also depend on the species of the atoms. For notational simplicity, we limit our discussion to systems of a single atomic species. However, KLIFF supports systems with arbitrary species.}
\begin{equation}
\mathcal{V} = \mathcal{V}(\bm r_1,\bm r_2,\dots,\bm r_N; \bm\theta) ,
\end{equation}
where $\bm r_1,\bm r_2,\dots,\bm r_N$ are the positions of a system of $N$
atoms, and $\bm\theta$ denotes a set of fitting parameters associated with the
IP functional form.
An IP must be invariant with respect to rigid-body translation and rotation, inversion of space, and permutation of chemically equivalent species according to the laws of physics \cite{tadmor2011modeling}.
These symmetry requirements are typically intrinsic to the functional form of the IP. For example, if an IP is expressed in terms of distances between atoms, it automatically satisfies the requirements of translation, rotation and inversion invariance.

%%%%%%%%%%%%%%%%%%%%%%%%%%%%%%%%%%%%%%%%
% PB potentials
%%%%%%%%%%%%%%%%%%%%%%%%%%%%%%%%%%%%%%%%
\subsection{Physics-based potentials}
\label{sec:pb:pot}

The functional form of a physics-based potential is carefully devised to model
the physics underlying the material system.
For example, as discussed above, the LJ potential
\cite{jones1924a,jones1924b,lennardjones1931} provides a good model for van der Waals
interactions in the noble gases, whereas for covalent systems more
complex forms are required, such as bond-order potentials \cite{brenner2002second,duin2001reaxff}.
Here, we briefly review the three-body Stillinger--Weber (SW) potential for silicon
\cite{stillinger1985computer} as an example.

\begin{figure}
\centering
\includegraphics[width=0.4\columnwidth]{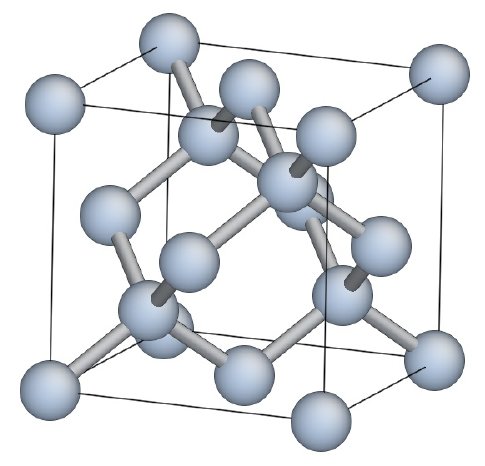}
\caption{Bulk silicon crystallizes in a diamond cubic crystal structure in
which each atom has four nearest neighbors forming the $sp^3$ hybridized
tetrahedral structure.}
\label{fig:diamond:si}
\end{figure}

The SW potential energy $\mathcal{V}$ of a system consisting of $N$ atoms
has the form,
\begin{equation}\label{eq:sw}
\mathcal{V} = \sum_{i=1}^N\sum_{j>i}^N \phi_2(r_{ij})
   + \sum_{i=1}^N\sum_{j \neq i}^N\sum_{\substack{k>j\\ k\neq i}}^N
     \phi_3(r_{ij}, r_{ik}, \beta_{jik}) ,
\end{equation}
where the two-body interaction takes the form
\begin{equation}
\phi_2(r_{ij})\label{eq:sw:phi2:ori}
   = \epsilon \hat{A} \left[B \left( \frac{r_{ij}}{\sigma}\right)^{-p} - \left(
   \frac{r_{ij}}{\sigma}\right)^{-q} \right]
\times \exp\left( \frac{1}{r_{ij}/\sigma - a} \right) ,
\end{equation}
and the three-body term is
\begin{align}\label{eq:sw:phi3:ori}
\phi_3(r_{ij}, r_{ik}, \beta_{jik})
   =& \epsilon \hat{\lambda} \left[ \cos\beta_{jik} - \cos\beta^0 \right]^2 \nonumber\\
   &\times
\exp\left(  \frac{\hat{\gamma}} {r_{ij}/\sigma - a}  +   \frac{\hat{\gamma}} {r_{ik}/\sigma - a}  \right) ,
\end{align}
in which $r_{ij} = \|\bm r_i - \bm r_j\|$ is the bond length between atoms $i$
and $j$, $\beta_{jik}$ is the bond angle formed by bonds $i$--$j$ and $i$--$k$
with the vertex at atom $i$.
%and $\beta^0$ is the corresponding predetermined reference angle.
The parameters are $\epsilon, \hat{A}, B, p, q, \sigma, a,
\hat{\lambda}, \hat{\gamma}$, and $\beta^0$.
The functional form is based on the lattice structure of bulk
silicon shown in \fref{fig:diamond:si}.
The two-body term (\eref{eq:sw:phi2:ori}) models bond stretching and
compression, and the three-body term (\eref{eq:sw:phi3:ori})
penalizes configurations away from the tetrahedral ground state
structure of silicon.

%%% diamond silicon structure

The cutoff distance in the SW potential is implicitly defined as $r^\text{cut}=a\sigma$.
This is not ideal from a potential fitting perspective.
When fitting an IP, it is typical to fix the cutoff distance, and
then adjust other parameters to minimize a loss function (discussed
later in \sref{sec:fitting}).
For the standard form of SW, both $a$ and $\sigma$ must be fixed to
set the cutoff. However, this adds an unnecessary constraint since two
parameters are fixed instead of just the cutoff. If instead only $a$
or $\sigma$ are fixed (or neither), then the cutoff will vary during the
fitting process. This can lead to failure of the optimization
due to discontinuity in the loss function when neighbors enter or leave the
cutoff sphere of an atom.
In addition to the cutoff problem, another issue with the SW form is that
$\epsilon$ is a redundant parameter that only scales the energy.

To avoid these pitfalls,
Eqs.\ \eqref{eq:sw:phi2:ori} and \eqref{eq:sw:phi3:ori}
are recast in a form in which all parameters are independent and the dependence on
the cutoff radius is made explicit \cite{wen2017sw}.
Let $A \coloneqq \epsilon\hat{A}$,
$\lambda \coloneqq  \epsilon\hat{\lambda}$,
$\gamma \coloneqq  \sigma\hat{\gamma}$,
and $r^\text{cut} \coloneqq a\sigma$, we have
\begin{subequations}
\begin{equation}\label{eq:sw:phi2}
\phi_2(r_{ij})
   = A \left[B \left( \frac{r_{ij}}{\sigma}\right)^{-p} - \left( \frac{r_{ij}}{\sigma}\right)^{-q} \right]
   \times \exp\left( \frac{\sigma}{r_{ij} - r^\text{cut}} \right) ,
\end{equation}
\begin{align}\label{eq:sw:phi3}
\phi_3(r_{ij}, r_{ik}, \beta_{jik})
   =& \lambda \left[ \cos\beta_{jik} - \cos\beta^0 \right]^2 \nonumber\\
   &\times \exp\left( \frac{\gamma} {r_{ij} - r^\text{cut}} + \frac{\gamma} {r_{ik} -
r^\text{cut}} \right).
\end{align}
\end{subequations}
The new parameters are $A, B, p, q, \sigma, \lambda, \gamma$ along with the
cutoff radius $r^\text{cut}$ and the equilibrium bond angles $\beta^0$.
The SW model implemented in KIM \cite{MD_335816936951_004}
takes the form of Eqs.\ \eqref{eq:sw:phi2}
and \eqref{eq:sw:phi3} instead of Eqs.\ \eqref{eq:sw:phi2:ori} and
\eqref{eq:sw:phi3:ori}.

%%%%%%%%%%%%%%%%%%%%%%%%%%%%%%%%%%%%%%%%
% ML potentials
%%%%%%%%%%%%%%%%%%%%%%%%%%%%%%%%%%%%%%%%
\subsection{Machine learning potentials}
\label{sec:ml:pot}

In contrast to physics-based potentials whose functional forms aim to capture the
physics underlying the material system, machine learning potentials employ
general-purpose regression models that interpolate across a dense training set
of first principles energies and forces. Similar to a physics-based potential,
a machine learning model returns the energy of an atom based on a finite
neighborhood of atoms in its vicinity.
Directly using the positions of an atom and its neighbors as input to the
machine learning potential is ill-advised since this would require the model to
learn the physical invariances of the IP
\cite{tadmor2011modeling,bartok2013representing},
significantly increasing the complexity of the model and required training data.
Instead, the atomic environment in terms of positions is
transformed to a suitable ``descriptor'' vector representation that identically
satisfies all invariances. For example two atomic environments that differ only
by a rigid-body rotation would yield the same descriptor vector.
Various descriptors have been developed to represent atomic
environments, including the Coulomb matrix \cite{rupp2012fast},
symmetry functions \cite{behler2007generalized,behler2011atom},
bispectrum \cite{bartok2010gaussian,bartok2013representing,thompson2015spectral},
many-body tensor \cite{huo2017unified}, and others \cite{langer2020representations}.
As an example, we briefly review the symmetry functions approach,
which is one of the earliest and most intuitive representations.
For a more detailed discussion, see for example Ref.~\cite{wen2019thesis}.

%%%%%%%%%%%%%%%%%%%%
% symmetry functions
%%%%%%%%%%%%%%%%%%%%
The symmetry functions \cite{behler2007generalized,behler2011atom} are
comprised of a set of two-body \emph{radial functions} and a set of
three-body \emph{angular functions}. Specifically,
the environment of atom $i$ is
characterized by three types of radial functions:
\begin{align}
\label{eq:g1}
    G_i^1 &= \sum_{j\neq i}  f_\text{c}(r_{ij}), \\
\label{eq:g2}
    G_i^2 &= \sum_{j\neq i} e^{-\alpha (r_{ij} - R_\text{s})^2} f_\text{c}(r_{ij}), \\
\label{eq:g3}
    G_i^3 &= \sum_{j\neq i} \cos(\kappa r_{ij})
    f_\text{c}(r_{ij}),
\end{align}
and two types of angular functions:
\begin{align}
\label{eq:g4}
    G_i^4 &= 2^{1-\zeta} \sum_{j\neq i} \sum_{\substack{k>j\\ k\neq i}}
    (1+\lambda \cos\beta_{jik})^\zeta
    e^{-\eta(r_{ij}^2 + r_{ik}^2 + r_{jk}^2)}
    f_\text{c}(r_{ij}) f_\text{c}(r_{ik}) f_\text{c}(r_{jk}), \\
\label{eq:g5}
%   G_i^5 &= 2^{1-\zeta} \sum_{j\neq i} \sum_{\substack{k>j\\ k\neq i}}
%   (1+\lambda \cos\beta_{jik})^\zeta
%   e^{-\eta(r_{ij}^2 + r_{ik}^2 + r_{jk}^2)}
%   f_\text{c}(r_{ij}) f_\text{c}(r_{ik}),
    G_i^5 &= 2^{1-\zeta} \sum_{j\neq i} \sum_{\substack{k>j\\ k\neq i}}
    (1+\lambda \cos\beta_{jik})^\zeta
    e^{-\eta(r_{ij}^2 + r_{ik}^2)}
    f_\text{c}(r_{ij}) f_\text{c}(r_{ik}),
\end{align}
where $r_{ij}$ and $\beta_{jik}$ are distance and angle as defined in
\sref{sec:pb:pot}, and
$\alpha, R_s, \kappa, \zeta, \lambda$, and  $\eta$ are hyperparameters.
The cutoff function $f_\text{c}$ is given by
\begin{equation}\label{eq:sym_fn:cutoff}
f_\text{c}(r) =
\begin{cases}
  \frac{1}{2} \left[\cos\left(\frac{\pi r}{r^\text{cut}}\right) + 1\right] &\ \text{for} \   r\leq r^\text{cut} \\
  0 &\ \text{for}\  r > r^\text{cut}
\end{cases},
\end{equation}
where $r^\text{cut}$ is the cutoff distance beyond which atoms do not
contribute to the local environment.

\correction{
The symmetry functions depend on both distances and angles, however since
angles can be expressed in terms of distances through the law of cosines,
the symmetry functions depend entirely on distances and are therefore
invariant with respect to translation,
rotation, and inversion of space \cite{tadmor2011modeling}.
}
The symmetry functions also satisfy the permutation symmetry requirement,
because they are constructed by summation over all bond lengths and bond
angles within the cutoff sphere and changing the summation order does not affect the results.
%They are obviously continuous and differentiable.
One can select all the symmetry functions $G_i^1 \dots G_i^5$ to describe the atomic environment or a subset. As an example, we select
one radial function and one angular function, $G_i^2$ and $G_i^4$.
The descriptor  vector is comprised of distinct $G_i^2$ and $G_i^4$ values
obtained for different choices of the hyperparameter sets \{$\alpha$, $R_s$\}
and \{$\lambda$, $\zeta$, $\eta$\}, respectively.
The length of the descriptor vector is then equal to the total number of
hyperparameter sets, $N_{G_i^2}+N_{G_i^4}$.
(See the supplementary material \cite{supplementary} for the hyperparameter sets for  $G_i^2$ and $G_i^4$ used in \sref{sec:parameterization}.)
%independent of the number of neighboring atoms.
%The symmetry functions are not complete because interactions of orders higher
%than two-body and three-body are ignored.

%%%%%%%%%%%%%%%%%%%%
% neural nets
%%%%%%%%%%%%%%%%%%%%

% nn struct
\begin{figure}
\centering
\includegraphics[width=1.0\columnwidth]{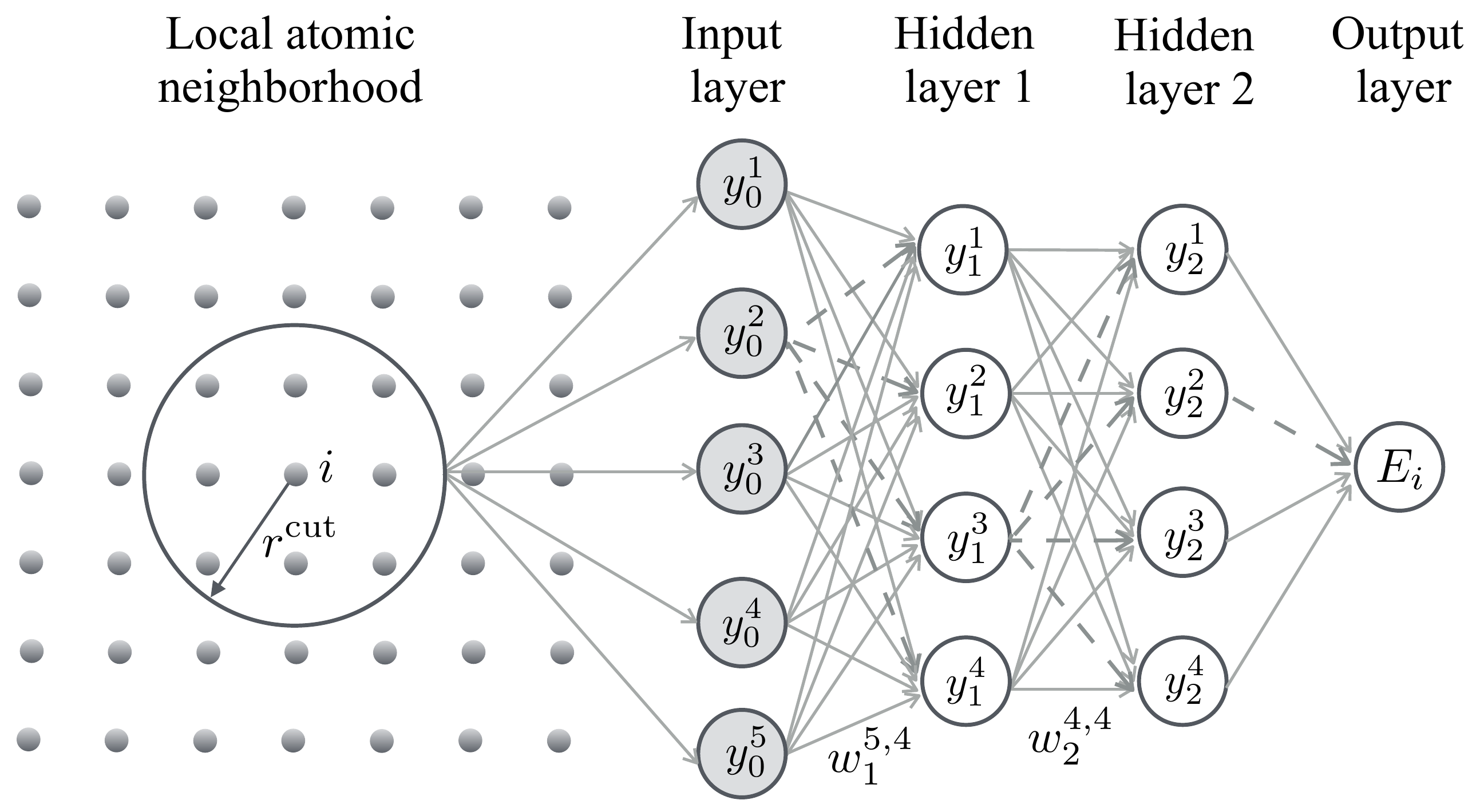}
\caption{Schematic representation of an NN potential to compute the atomic
energy $E_i$.
The NN consists of an input layer, two hidden layers and an output layer.
The local atomic neighborhood information of atom $i$ (all atoms within a sphere of radius $r^\text{cut}$
around atom $i$)
is transformed to descriptor vector
with components $y_0^j$ ($j=1,2,\dots$) that serves as the input to the NN.
Each arrow connecting two nodes between adjacent NN layers represents a weight.
The fully-connected NN becomes a dropout NN when some connections are cut
(e.g.\ removing the dashed arrows).
Biases and activation function are not shown in this plot.
See text for explanation of the variables.
}
\label{fig:nn:struct}
\end{figure}

Many machine learning regression methods are suitable for constructing IPs including
parametric linear regression and neural network (NN) models,
nonparametric kernel ridge regression and Gaussian process models, and others \cite{langer2020representations}.
%IPs built using these methods have been shown successful in reproducing
%first-principles energies and forces and are widely used to study a variety of
%materials problems.
%A detailed discussion of the regression methods applied to IPs can be found in Ref.~\cite{wen2019thesis}.
Here, we discuss the NN model.
In an NN potential, the total potential energy of a configuration consisting of
$N$ atoms is decomposed into the contributions of individual atoms
\begin{equation}\label{eq:eng:decompose}
\mathcal{V} = \sum_{i=1}^N E_i,
\end{equation}
where $E_i$ is the energy of atom $i$, represented by an NN as shown in
\fref{fig:nn:struct}.
The NN returns the energy $E_i$ based on
the positions of atom $i$ and its neighbors up to a cutoff distance
$r^\text{cut}$.
%The use of a cutoff significantly reduces the computational cost by restricting
%the dependence of an atom's energy to its local environment.
The values $y_0^1, y_0^2, \dots$ in the input layer are the components of the
descriptor.
Between the input layer and the energy output layer are so-called ``hidden''
layers that add complexity to the NN\@.
In a fully-connected NN, each node in a hidden layer is connected to all the
nodes in the previous layer and in the following layer.
The value of node $n$ in layer $m$ is\footnote{The input layer and the output layer are indexed as the zeroth layer and third layer, respectively.}
\begin{equation}\label{eq:nn:connect}
y_m^n = \act \left(\sum_{n'} y_{m-1}^{n'} w_{m}^{n',n} + b_m^n \right),
\end{equation}
where $w_m^{n',n}$ is the weight
that connects node $n'$ in layer $m-1$ and node $n$ in layer $m$, $b_m^n$ is the
bias applied to node $n$ of layer $m$, and $\act$ is an activation function
(e.g.\ hyperbolic tangent) that introduces nonlinearity into the NN\@.
In a more compact way, \eref{eq:nn:connect} can be written as $\bm y_m =
\act(\bm y_{m-1} \bm W_m + \bm b_m)$, where $\bm y_m$ is a row vector of the node values in
layer $m$, $\bm W_m$ is a weight matrix, and $\bm b_m$ is a row vector of the
biases.
For example, $\bm y_1$ and $\bm b_1$ are row vectors each with 4 elements and
$\bm W_1$ is a $5\times 4$ matrix for the NN shown in \fref{fig:nn:struct}.
Consequently, the atomic energy $E_i$ represented in \fref{fig:nn:struct} can be
expressed as
\begin{equation} \label{eq:nn:compact}
  E_i = \act[\act[\bm y_0 \bm W_1 + \bm b_1]\bm W_2 +  \bm b_2]\bm W_3 + \bm b_3.
\end{equation}
The weights and biases are the fitting parameters in an NN potential:
$\bm\theta = \{\bm W_1, \bm W_2,\dots,\bm W_L, \bm b_1, \bm b_2, \dots, \bm b_L \}$,
 where $L$ is the number of layers (hidden and output).

%%%%%%%%%%%%%%%%%%%%%%%%%%%%%%%%%%%%%%%%
% fitting
%%%%%%%%%%%%%%%%%%%%%%%%%%%%%%%%%%%%%%%%
\subsection{Parameterization}
\label{sec:fitting}

Once an IP functional form is selected, the parameters must be determined.
This is typically framed as a least-squares
minimization problem where the IP parameters are adjusted to best match
a training set of reference data obtained from experiments and/or
first-principles calculations.
For a training set of $M$ configurations, the difference between
the predictions of the IP and the reference data is quantified by a
\emph{loss} function defined as
\begin{equation}\label{eq:loss}
\mathcal{L}(\bm\theta)
= \frac{1}{2}\sum_{m=1}^M w^\text{e}_m \left[E(\bm R_m; \bm\theta) - \hat{E}_m \right]^2
+ \frac{1}{2}\sum_{m=1}^M  w^\text{f}_m \|\bm f(\bm R_m; \bm\theta) - \hat{\bm f}_m \|^2,
\end{equation}
where $E(\bm R_m; \bm\theta) \in \mathbb{R}$ and
$\bm f (\bm R_m; \bm\theta)
= -\left.(\partial E/\partial \bm{R})\right|_{\bm R_m}
\in \mathbb{R}^{3N_m} $
are the energy and forces in configuration $m$ obtained from an IP,
$\hat{E}_m$ and $\hat{\bm f}_m$ are the corresponding reference energy
and forces for configuration $m$ in the training set,
with $\bm R_m \in \mathbb{R}^{3N_m}$ the concatenated coordinates
of all atoms in configuration $m$ and $N_m$ the number of atoms in configuration $m$.
The weights $w^\text{e}_m$ and $w^\text{f}_m$ are typically chosen to be inversely
proportional to $(N_m)^2$, so that each configuration has an equal contribution to the loss function $ \mathcal{L}(\bm\theta)$.
This prevents configurations with more atoms from dominating the optimization.
For energy in units of eV and forces in units of eV/\AA, these weights have
units of eV$^{-2}$ and (eV/\AA)$^{-2}$, respectively.
Here, we only use energy and forces to construct the loss function, but in
principle one can fit any physical property, such as the equilibrium lattice
constants and elastic moduli of a ground state crystal structure.
The objective then is to minimize the loss function in \eref{eq:loss} with respect to $\bm\theta$ to obtain the optimal
set of IP parameters.

\correction{
Simply minimizing \eref{eq:loss} can lead to overfitting and thus low transferability of an IP.
This is especially true for machine learning IPs due to the lack of physics in their functional forms and the large parameter space.
Various techniques have been proposed to overcome this problem.
One approach is to add regularization terms to the loss function to prevent
overly complex results, for example an $L_2$ term of the form
$\lambda \|\bm\theta\|^2$ can be added, where $\lambda$ is a hyperparameter that
determines the regularization weight.
Another widely used approach is early stopping \cite{prechelt1998automatic},
where model performance is monitored on a validation set and fitting is
terminated when accuracy begins to degrade.
There are also regularization techniques that are specific to certain types
of models. For example, the dropout method \cite{hinton2012improving,srivastava2014dropout}
can be applied to NN potentials (see \sref{sec:uncertainty} for more on dropout).
}

%To get the parameters, minimization of the loss in \eref{eq:loss} is necessary
%for almost all parametric models.
%One exception is linear regression, where the parameterizes can be obtained physics-basedally.
%Although we can still minimize the loss to get the parameters, a physics-based
%solution exists and it can be obtained simply by matrix inversion.
%For some nonparametric models, minimization of the loss is also not necessary.
%KLIFF supports the parameterization of these models that do not need to carry
%out minimization.
%For simplicity and brevity, we focus on models that require a minimization in
%the following discussions.

%%%%%%%%%%%%%%%%%%%%%%%%%%%%%%%%%%%%%%%%%%%%%%%%%%%%%%%%%%%%%%%%%%%%%%%%%%%%%%%%
% features
%%%%%%%%%%%%%%%%%%%%%%%%%%%%%%%%%%%%%%%%%%%%%%%%%%%%%%%%%%%%%%%%%%%%%%%%%%%%%%%%
\section{Features and capabilities of KLIFF}
\label{sec:features}

A variety of software packages have been developed to develop IPs, including
potfit \cite{brommer2007potfit,brommer2015classical},
%ForceFit \cite{waldher2010forcefit},
%ForceBalance \cite{wang2014building},
\ae net \cite{artrith2016implementation},
Amp \cite{khorshidi2016amp},
\correction{
aPIP \cite{allen2021atomic},
}
atomicrex \cite{stukowski2017atomicrex},
DeePMD-kit \cite{wang2018deepmd},
\correction{
GAP \cite{gap,bartok2010gaussian},
}
MAISE \cite{hajinazar2021maise},
MLIP \cite{novikov2020mlip},
%PES-Learn \cite{Abbott2019},
\correction{
PACE \cite{lysogorskiy2021performant},
}
PANNA \cite{lot2020panna},
PyXtal\_FF \cite{yanxon2020pyxtal_ff},
\correction{
RuNNer \cite{runner,behler2007generalized},
}
SIMPLE-NN \cite{lee2019simple},
among others.
KLIFF shares many features with these packages, but is also distinguished
by some capabilities described in this section that address the problems discussed in \sref{sec:intro}.

%%%%%%%%%%%%%%%%%%%%%%%%%%%%%%%%%%%%%%%%
% KIM
%%%%%%%%%%%%%%%%%%%%%%%%%%%%%%%%%%%%%%%%
\subsection{Integration with KIM}
\label{sec:integ:kim}

As indicated by the name, KLIFF is deeply integrated with the KIM ecosystem.
%Computer implementations of IP are archived in OpenKIM, verified
%for coding integrity, and tested by computing their predictions for a variety of
%material properties.
%Models conforming to the KIM API work seamlessly with major
%simulation packages
%that have adopted the KIM API standard.
%KLIFF integrates with KIM in a number of ways.
(We note that the Potfit IP fitting
framework is also compatible with KIM \cite{wen2017potfit}.)

First, KLIFF supports the training of IPs archived within
the OpenKIM repository.
An IP is called a \emph{model} in KIM nomenclature, and
a KIM \emph{portable model} is an independent computer implementation of an IP that
conforms to the KIM API portable model interface (PMI) standard.\footnote{KIM also supports a second
type of model called a \emph{simulator model}.
While a portable model will work seamlessly with any simulation package that
supports the KIM API/PMI standard, a simulator model only specifies how to setup
and run a model that is implemented as an integrated part of a specific
simulation package. KLIFF supports the fitting of portable models.}
In practice portable models consist of a ``model driver,'' which implements an IP class
(e.g. the embedded atom method (EAM) form) and a parameter set for a specific set of
species.
%It can be a stand-alone model or a model driver that reads in different
%parameter files to define different models.
All content in the OpenKIM repository is archived subject to
strict versioning and provenance control with digital object identifiers (DOIs) assigned.
This makes it possible to access the exact IP used in a publication at a later
date to reproduce the calculations or to conduct further fitting.
A large number of physics-based and machine learning IPs are implemented as portable models and archived in the OpenKIM repository.
%A large number of IPs are implemented as portable models and archived in the OpenKIM repository,
%such as
%the SW potential \cite{MD_335816936951_004, stillinger1985computer} discussed in
%\sref{sec:pb:pot},
%the Tersoff potential \cite{MD_077075034781_003, MD_077075034781_003a, MD_077075034781_003b, MD_077075034781_003c},
%the environment-dependent interatomic potential (EDIP) \cite{MD_506186535567_002, MD_506186535567_002a, MD_506186535567_002b, MD_506186535567_002c},
%the embedded-atom method (EAM) \cite{MD_120291908751_005, daw1984embedded, daw1993embedded},
%modified embedded-atom method (MEAM) \cite{MD_249792265679_001, %MD_249792265679_001b, MD_249792265679_001c},
%to name just a few.
These models are subjected to an editorial review process by the
KIM Editor on acceptance to ensure quality control.
Users of KLIFF can employ these models directly without
having to implement them with significant savings in time and potential errors.

Second, IPs trained with KLIFF can be easily tested via OpenKIM\@.
KLIFF can automatically generate models that are compatible with the KIM API,
thus allowing a trained IP to run against KIM verification checks (VCs) and KIM
tests \cite{karls:bierbaum:2020}.
As noted in \sref{sec:intro}, KIM VCs are programs that explore the integrity of an IP
implementation.
They check for programming errors (e.g.\ memory leak
\cite{VC_561022993723_001}), failures to satisfy required behaviors (e.g.\
inversion \cite{VC_021653764022_001} and permutation \cite{VC_903502816694_001}
symmetries), and determine general characteristics of the IP functional form
(e.g.\ are the forces returned by the model consistent with those obtained
through numerical differentiation of the energy \cite{VC_710586816390_002}).
As opposed to KIM VCs, KIM tests check the accuracy of an IP by computing a
variety of physical properties of interest to researchers, such as the stacking
fault energy \cite{TD_228501831190_001}, elastic moduli
\cite{TD_011862047401_005}, and linear thermal expansion coefficient
\cite{TD_522633393614_000} to name a few.
The information provided by KIM VCs and KIM tests can save researchers a great
deal of time by identifying limitations of an IP that can lead to subtle
problems in simulations (e.g.\ poor convergence during energy minimization due
to incorrect or discontinuous forces), and assisting in the selection of IPs by considering its predictions for relevant physical properties.

Third, IPs trained with KLIFF can be deployed via KIM\@.
%Up to the writing of the paper, most IP development papers only report the functional form of
%the IPs and the associated parameters, without mentioning or providing any
%computer implementation.
%After obtaining a satisfied IP, IP developers either lack the interest or
%expertise to make the IP implementation publicly available by transplanting it
%to a molecular simulation code.
%Even if an IP developer is willing to go through this time-consuming and
%error-prone implementation transplanting process, the IP will end up in only one
%or two simulation packages that the IP developer thinks important.
%If a user happens to use the same simulation package, he/she will find
%himself/herself lucky enough to have access to the IP\@.
%Otherwise, users need to implement the IP themselves in the simulation package they
%want to use or have to wait until someone else to implement it.
Traditionally, most IP development papers only report the functional form of
the IPs and the associated parameters, without mentioning or providing a
computer implementation.
Recently developed machine learning potentials typically do provide computer implementations,
but these are often standalone codes that cannot be used in major molecular
simulation packages.
This creates a significant barrier for the universal usability of IPs.
By providing portable implementations, KIM addresses this issue, as
well as enabling reproducibility.\footnote{In some cases the same parameter
file can lead to different results when used with different implementations
of an IP, either in newer versions of the same code, or in different
simulation packages. For example, see Ref.~\cite{wen2015interpolation}
for a discussion of this effect for tabulated EAM potentials.}
As mentioned above, KLIFF can automatically create IP models that are compatible
with the KIM API, which enables the IP to work seamlessly with
any KIM-compliant simulation package including
ASE \cite{larsen2017atomic,ase}, DL\_POLY \cite{smith1996,dlpoly},
GULP \cite{gale1997gulp,gulp}, LAMMPS \cite{plimpton1995fast,lammps2021,lammps}, and QC \cite{tadmor1996,qc}.
The final production IP can also be contributed to the OpenKIM repository for
deployment as source and binary packages for major Mac, Linux and Windows
platforms.

%%%%%%%%%%%%%%%%%%%%%%%%%%%%%%%%%%%%%%%%
% uncertainty
%%%%%%%%%%%%%%%%%%%%%%%%%%%%%%%%%%%%%%%%
\subsection{Uncertainty analysis}
\label{sec:uncertainty}

Historically, molecular simulation with IPs has been primarily viewed
as a tool limited to providing qualitative insight.
A key reason is that such simulations include multiple sources of uncertainty
that are difficult to quantify, with the result that predictions obtained
from the simulation typically lack robust confidence intervals \cite{messerly2017uncertainty}.
A major source of uncertainty originates from the IPs themselves,
since these are empirical functional forms fitted to experimental results
and/or first-principles calculations.
To make molecular simulation with IPs more reliable, it is imperative to
quantify the intrinsic uncertainty of the IP and propagate it to
the simulation results.
This is an area that has not received much attention in the past.
% and thus most empirical IPs lack the ability to conduct uncertainty quantification.
To address this limitation, KLIFF provides functionality that enables uncertainty analysis of IPs.

\begin{figure}
\centering
\includegraphics[width=0.8\columnwidth]{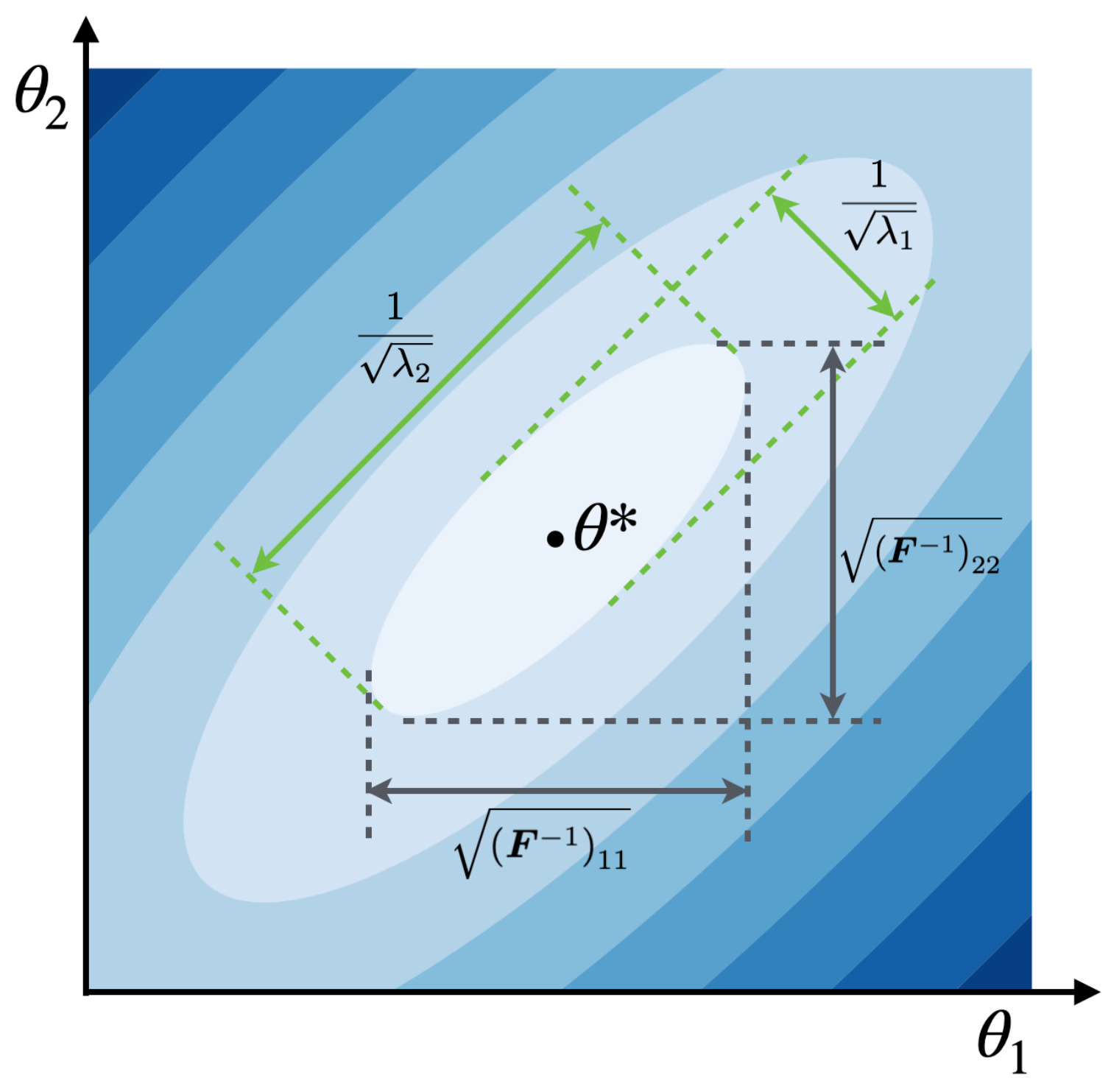}
\caption{Schematic representation of the cost contours in the vicinity of the optimal
parameters $\bm\theta^*$ of an IP with two parameters $\theta_1$ and $\theta_2$.
The aspect ratio of the contours is determined by the eigenvalues $\lambda_1$ and $\lambda_2$
of the FIM.
The diagonal elements of the inverse FIM $(\bm F^{-1})_{11}$ and $(\bm F^{-1})_{22}$
provide a lower bound on the variance of the parameters $\theta_1$ and $\theta_2$, respectively.
}
\label{fig:fim:contour}
\end{figure}

As mentioned in \sref{sec:intro},
IPs are often ``sloppy'' \cite{waterfall:casey:2006, wen2017sw, kurniawan2021bayesian}
in that their predictions are insensitive to certain combinations
of the parameters.
%The Fisher information matrix (FIM) provides a way to measure the amount of
%information that observable random variables (the training set here) carries
%about unknown parameters (IP parameters here) of a distribution\footnote{The
%least-squares minimization problem (with loss function in \eref{eq:loss}) of
%fitting an IP can be reformulated as a maximum likelihood problem with Gaussian
%distribution.} that models the observable random variables.
This behavior can be quantified using the notion of a Fisher information
matrix (FIM). The FIM provides a measure for the information in the training
set on the parameters, which leads to an estimate for the precision with which
the parameters can be determined \cite{wen2017sw}.
For example, for the case where the loss function contains only forces
(i.e.\ $w_m^\text{e}=0$ in \eref{eq:loss}), the FIM can be
written as \cite{wen2017sw, kurniawan2021bayesian}:
\begin{equation}\label{eq:fim}
\bm F(\bm\theta) \propto \sum_{m=1}^M
\left(\frac{\partial\bm f_m} {\partial\bm\theta}\right)^\text{T}
\left(\frac{\partial\bm f_m} {\partial\bm\theta}\right) ,
\end{equation}
where $\bm f_m \in \mathbb{R}^{3N_m}$ are the forces on atoms of configuration
$m$ ($N_m$ is the number of atoms in configuration $m$), \correction{$M$ is the number of configurations in the training set,} and the
superscript $\rm T$ denotes matrix transpose.
The diagonal elements of the inverse FIM provide lower bounds on the variance of
the parameters, known as the Cram\'er--Rao bound \cite{cramer11mathematical},
\begin{equation}\label{eq:cr:bound}
\text{Var}[\theta_i] \geq \left(\bm{F}^{-1} \right)_{ii}.
\end{equation}
The larger a diagonal element of the inverse FIM,
the larger the lower bound on the variance for the corresponding parameter,
indicating that the parameter is less well determined.
As an illustrative example, we plot in \fref{fig:fim:contour} a schematic representation of the contours
of the cost function in \eref{eq:loss} for an IP with two parameters.
Here, the two diagonal components of the inverse FIM $(\bm F^{-1})_{11}$ and $(\bm F^{-1})_{22}$
are nearly of the same magnitude, indicating that the two parameters $\theta_1$ and $\theta_2$ are
equally determined in the fitting.
If this were not the case and a diagonal component of the inverse FIM
was much larger (an order of magnitude or more) than the others,
then the IP parameter associated with this component is poorly determined.
To address this, one could attempt to modify the IP functional form as
discussed in \sref{sec:intro} and shown in \fref{fig:ipdev:flowchart}.
The FIM also provides an upper bound on the uncertainty in a physical quantity of interest (QOI)
due to small variations in IP parameters.
A detailed discussion of such an analysis for the thickness of monolayer MoS$_2$ can be found in  Ref.~\cite{wen2017sw}.
The FIM in \eref{eq:fim} is implemented in KLIFF as an \verb|Analyzer| (discussed in \sref{sec:implementation}) using numerical differentiation.

% fim contour

%The FIM also provides an upper bound on the uncertainty in any observable
%predicted by the IP that is obtained by averaging with respect to an
%equilibrium distribution in phase space due to infinitesemial variations in the parameters.
%The bound is a tighter version of the Csisz\'ar--Kullback--Pinsker inequality
%\cite{tsourtis2015parametric}, having the form \cite{dupuis2016path}:
%\begin{equation} \label{eq:ckp:bound}
%\left\vert \mathbb{E}_{\bm\theta+\Delta\bm\theta}[P] - \mathbb{E}_{\bm\theta}[P]
%\right\vert
%\leq \epsilon \sqrt{\text{Var}_{\bm{\theta}}[P]}  \sqrt{\Delta\bm\theta^\text{T}
%\bm F(\bm\theta)  \Delta\bm\theta},
%\end{equation}
%where $\Delta\bm\theta$ is a perturbation in the IP parameters, and
%$\mathbb{E}_{\bm\theta}[P]$ and $\text{Var}_{\bm{\theta}}[P]$ are the
%expectation and variance of the observable $P$.
%KLIFF includes functionality to use the IP's FIM to analyze the sensitivity
%in IP parameters and the uncertainty of an observable relative to
% parameter perturbation.
%See Ref.~\cite{wen2017sw} for an example of such an analysis.

The FIM analysis is well suited for physics-based potentials, which have dozens of parameters
and each parameter plays a vital role.
However, machine learning potentials are typically over-parameterized and the influence
of a single parameter on the model performance is not large.
Instead of parameter uncertainty, it is more important and useful to analyze the prediction
uncertainty of a QOI (e.g.\ elastic moduli).
A simple yet powerful approach to obtaining the QOI uncertainty is to construct an
\emph{ensemble} of IPs instead of a single best fit model.
This can be done by either training different IPs using different initial guesses for the
parameters or using different subsets of the training data.
At the prediction stage, each individual model in the ensemble is applied to compute the QOI $P$. The average
\begin{equation}
\bar P = \frac{1}{N_P}\sum_{i=1}^{N_P} P_i
\end{equation}
is then used as the predictive mean for the QOI,
and the standard deviation
\begin{equation}
\text{Std}[P] = \sqrt{\frac{1}{N_P-1} \sum_{i=1}^{N_P} (P_i - \bar P)^2}
\end{equation}
as the uncertainty.
%The ensemble approach can be applied to any type of model, but it is computationally expensive
%since multiple models have to be trained to form the ensemble.
\correction{
The ensemble approach can be applied to any type of model, either physics-based \cite{frederiksen2004bayesian,longbottom2019uncertainty} or machine learning potentials  \cite{artrith2012high,peterson2017addressing,zhang2019active,jeong2020efficient}.
Although straightforward to train, it is computationally expensive since multiple models have to be trained to form the ensemble.
}
For NN potentials, there is an alternative that is computationally less costly and performs
equally well to the ensemble approach \cite{wen2020uncertainty}.
By removing some connections between layers (e.g.\ removing the dashed arrows
for the NN shown in \fref{fig:nn:struct}), a fully-connected NN is changed into a
dropout NN \cite{hinton2012improving, srivastava2014dropout}.
It has been shown that training an NN with dropout (i.e.\ dropping different connections
at each training step)
approximates a Bayesian NN \cite{gal2016dropout, gal2016uncertainty}.
Consequently, a dropout NN possesses all the properties of a probabilistic
Bayesian model, from which uncertainty information can be extracted.
For dropout NN potentials \cite{wen2020uncertainty}, only one model needs to be trained at
the training stage.
At the prediction stage, it is essentially an ensemble model and can be used in a similar fashion:
conduct multiple stochastic forward passes through the dropout NN (each time drop different connections)
to obtain multiple samples of the QOI and then compute the average and standard deviation.
KLIFF supports the training of both ensemble and dropout NN potentials. The associated
KIM DUNN model driver \cite{MD_292677547454_000} allows molecular simulation codes to work with individual members in the ensemble and perform uncertainty quantification.

%In practice, to make predictions for a new QOI using a dropout NN
%potential, we only need to do multiple stochastic forward passes
%through the dropout NN (each time drop different connections) to get multiple
%samples of the output $P_1, P_2,\dots, P_N$ at the prediction stage.
%The average
%\begin{equation}
%\bar P = \frac{1}{N}\sum_{i=1}^{N} P_i
%\end{equation}
%and the variance
%\begin{equation}
%\text{Var}[P] = \sqrt{\frac{1}{N-1} \sum_{i=1}^{N} (P_i - \bar P)^2}
%\end{equation}
%of these samples can then be computed as the predictive mean and uncertainty,
%respectively \cite{wen2020uncertainty}.
%KLIFF supports the training of dropout NN potentials to conduct uncertainty
%quantification in molecular simulations.

%%%%%%%%%%%%%%%%%%%%%%%%%%%%%%%%%%%%%%%%
% scipy and pytorch
%%%%%%%%%%%%%%%%%%%%%%%%%%%%%%%%%%%%%%%%
%\subsection{State-of-the-art tool sets}
\subsection{A wide range of support}
\label{sec:wide:support}

By conforming to the KIM API, KLIFF supports a wide range of IPs available
through OpenKIM.  At the time of this writing, the OpenKIM repository contains
35 model drivers, including widely used physics-based potentials such as
Stillinger--Weber (SW) \cite{MD_335816936951_004, stillinger1985computer},
Tersoff \cite{MD_077075034781_003, MD_077075034781_003a, MD_077075034781_003b, MD_077075034781_003c},
EDIP \cite{MD_506186535567_002, MD_506186535567_002a, MD_506186535567_002b, MD_506186535567_002c},
and EAM \cite{MD_120291908751_005, daw1984embedded, daw1993embedded}
potentials among others.
For machine learning potentials, KLIFF currently supports the symmetry functions \cite{behler2007generalized,behler2011atom} and
bispectrum \cite{bartok2010gaussian,bartok2013representing} atomic environment descriptors.
Interfacing with other descriptor libraries, such as DScribe \cite{himanen2020dscribe},
is being explored.
For machine learning regression algorithms, KLIFF has its own implementation of simple algorithms
(e.g.\ linear regression)
%Deep learning with NN is a fast growing area where new techniques are
%proposed every few months.
%To avoid reinvent the wheels,
and takes advantage of PyTorch \cite{pytorch}
%(an open-source deep learning platform that provides a seamless path from research prototyping to production deployment)
to build and train NN potentials.
The NN model in KLIFF wraps PyTorch so that the user
interface appears the same as other models in KLIFF, % (more on this in \sref{sec:uniform}),
but still retains the flexibility of PyTorch to
create customizable NN structures and train with state-of-the-art deep
learning techniques available through this package.

KLIFF provides an interface to many widely-used minimization algorithms for model training.
As discussed in \sref{sec:fitting}, the IP parameters are obtained by minimizing
a loss function that quantifies the difference between IP predictions and the
training set.
The optimizer directly determines the values of the parameters and thus the quality
of the IP\@.
It is impossible to make a general statement about which optimizer is best, since this
is problem-dependent, but some optimizers (e.g.\ the L-BFGS-B algorithm
\cite{zhu1997algorithm}) tend to work well for a wide range of problems.
KLIFF supports the optimization algorithms in SciPy \cite{scipy} and
PyTorch \cite{pytorch}.
The \verb|minimize| module of \verb|scipy.optimize| provides a large number of
general-purpose minimization algorithms,
%such as the conjugate gradient (CG) method \cite{hestenes1952methods}, the
%Powell method \cite{powell1965method}, the L-BFGS-B methods \cite{zhu1997algorithm};
and the \verb|least_squares| module of \verb|scipy.optimize| provides algorithms
specific for nonlinear least-squares minimization problems.
%with a loss function of the form in \eref{eq:loss} such as the Levenberg--Marquardt (LM)
%method \cite{levenberg1944algorithm,marquardt1963algorithm}.
%such as the trust region reflective method \cite{branch1999subspace}, the
%dogleg method with rectangular trust regions \cite{dennis1996numerical} and the
%Levenberg--Marquardt (LM) method \cite{levenberg1944algorithm,marquardt1963algorithm}.
The optimizers in PyTorch are targeted for training NN models, including the
stochastic gradient descent (SGD) method \cite{robbins1951stochastic,
kiefer1952stochastic} and its variants such as the Adam method \cite{kingma2014adam}.
In addition, KLIFF also supports the geodesic Levenberg--Marquardt (LM) algorithm
\cite{transtrum2011geometry,transtrum2012geodesic,transtrum2012improvements},
which has been shown to work well for ``sloppy'' IPs whose
predictions are insensitive to certain parameters or certain combinations of
parameters.
% (See \cite{wen2017potfit} for a comparison of the use of the
% geodesic LM method and other methods to fit the EDIP potential for silicon.)

%%%%%%%%%%%%%%%%%%%%%%%%%%%%%%%%%%%%%%%%
% modularity
%%%%%%%%%%%%%%%%%%%%%%%%%%%%%%%%%%%%%%%%
\subsection{Uniformity, modularity, and extensibility}
\label{sec:uniform}

KLIFF is designed to be as uniform, modular, and extensible as possible.
It is implemented using an object-oriented programming (OOP) paradigm and
provides a pure Python user interface.
All the atomic environment descriptors, models, calculators, analyzers, etc.\ are
subclassed from individual superclasses.
A subclass only provides or modifies specific implementations of superclass
methods when necessary, %without changing their names and arguments,
guaranteeing a uniform interface across subclasses.
As mentioned in \sref{sec:wide:support}, KLIFF takes advantage of the
optimization algorithms in SciPy \cite{scipy} and PyTorch \cite{pytorch}, as well as
the geodesic LM algorithm
to train models when minimization of a loss function is needed.
Although vanilla SciPy, PyTorch, and LM have different APIs to call the
optimization algorithms, KLIFF provides a unified interface that wraps them under the hood.

%The task to train an IP can be divided into several subtasks.
%For example, to train a machine learning IP we typically:
%(1) select a descriptor to transform the atomic environments to vector representations;
%(2) build a regression model that takes the vector representations as input to
%calculate a set of predictions (e.g.\ energy and forces);
%(3) construct a loss function based on the predictions of the model and the corresponding
%reference data in the training set and then minimize the loss function to obtain
%the optimal parameter set; and
%(4) analyze the quality of the trained model.
%Using the OOP paradigm, each atomic environment descriptor,
%regression model, loss function, and analyzer are constructed as an individual
%module in such a way that any descriptor should work seamlessly with any
%regression model, any regression model should work with any loss function and so
%forth.

Extending KLIFF is straightforward.
New descriptors, models, calculators, loss functions, optimization algorithms,
analyzers, etc.\ can be seamlessly added to existing modules in KLIFF\@.
%For example, a new physics-based IP can be implemented either as a
%KIM portable model or a KLIFF model.
%The benefits of KIM models are discussed in \sref{sec:integ:kim};
%however, this may be a bit burdensome in cases where fast research prototyping
%of new functional forms of an IP is preferred, because one needs to get
%familiar with the KIM API and code in compiled languages such as C, C++, and
%Fortran.
%To this end, KLIFF allows the creation of new models within its own
%framework using pure Python and provides all sorts of utilities (e.g.\ neighbor
%list functions) to aid the implementation of an IP\@.
%With these utilities, typically, the only thing left is to use Python to code up
%the functional form of the IP.
For example, a new physics-based potential can be easily implemented by subclassing the
KLIFF ``Model'' class, specifying the IP parameters, and then using Python to code
the functional form of the IP.
\correction{
As a concrete example, we provide a Python code demonstrating how to implement the
Lennard-Jones potential in the supplementary material \cite{supplementary}.
}
Other parts such as periodic boundary conditions handling are dealt with by KLIFF.
The newly created model can then be used for training with any
loss function and optimization algorithms that are available in KLIFF.
To gain the benefits of KIM models discussed in \sref{sec:integ:kim},
it is currently necessary to implement the IP as a separate code
conforming to the KIM API. Future plans include the development of a
general KIM model driver that will directly work with KLIFF IPs
stored in a portable format.

%%%%%%%%%%%%%%%%%%%%%%%%%%%%%%%%%%%%%%%%
% parallel
%%%%%%%%%%%%%%%%%%%%%%%%%%%%%%%%%%%%%%%%
\subsection{Data parallelization}
\label{sec:parallel}

% parallel
\begin{figure}
\centering
\includegraphics[width=1.\columnwidth]{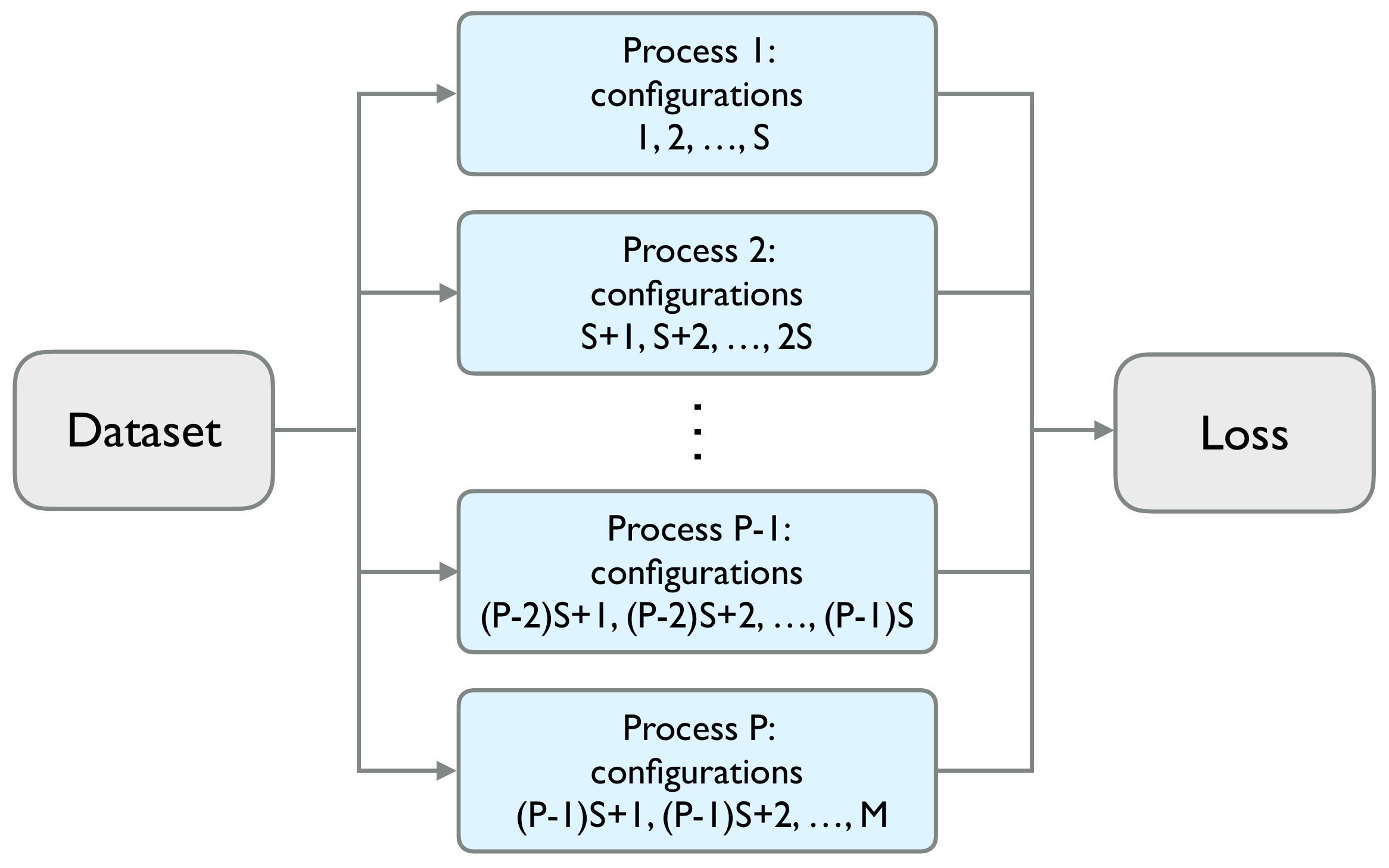}
\caption{Data parallelization scheme used by KLIFF.
S is the number of configurations assigned to each process, and M is the total number of configurations.}
\label{fig:data:parallel}
\end{figure}

Computationally intensive KLIFF components, such as neighbor list generation and
descriptor calculation, are internally implemented in C++.
Even with this, the computational requirements can become quite demanding
as the size of the training set increases.
Fortunately, evaluation of the loss function \eref{eq:loss} can be easily
divided into independent sub-problems allowing for easy parallelization.
KLIFF adopts the parallelization over data scheme illustrated in
\fref{fig:data:parallel}.
Atomic configurations in the dataset are distributed to different processes.
%\footnote{Each process may not get the same number of configurations (when the configurations do not have the same number of atoms),
%but instead the total number of atoms of the configurations distributed to each process is approximately the same.
%This balances the load of each process since the computational cost of IPs
%scales linearly with the number of atoms.}
Each process computes the sub-loss according to \eref{eq:loss} for the
configurations assigned to it, and the total loss is then obtained as the sum of
the sub-losses from all the processes.
%Besides the evaluation of the loss function, other tasks operating on the
%dataset such as generating the neighbor lists and computing the descriptor
%values are also parallelized in the same way.
KLIFF supports both OpenMP-style parallelism for shared-memory architectures,
and MPI-style parallelism typical of high-performance computing clusters
composed of multiple standalone machines connected by a network.
%Internally, we implement the data parallelization using both the OpenMP-style
%\verb|multiprocessing| module from Python and the MPI-style \verb|mpi4py|
%\cite{mpi4py} third-party package.
%These two are mutually exclusive.
%The former can only work on desktop machines, while the latter works on both
%desktop machines and HPC clusters.

%%%%%%%%%%%%%%%%%%%%%%%%%%%%%%%%%%%%%%%%%%%%%%%%%%%%%%%%%%%%%%%%%%%%%%%%%%%%%%%%
% implementation
%%%%%%%%%%%%%%%%%%%%%%%%%%%%%%%%%%%%%%%%%%%%%%%%%%%%%%%%%%%%%%%%%%%%%%%%%%%%%%%%
\section{Implementation details: the KLIFF code}
\label{sec:implementation}

\begin{figure}
\centering
\includegraphics[width=0.9\columnwidth]{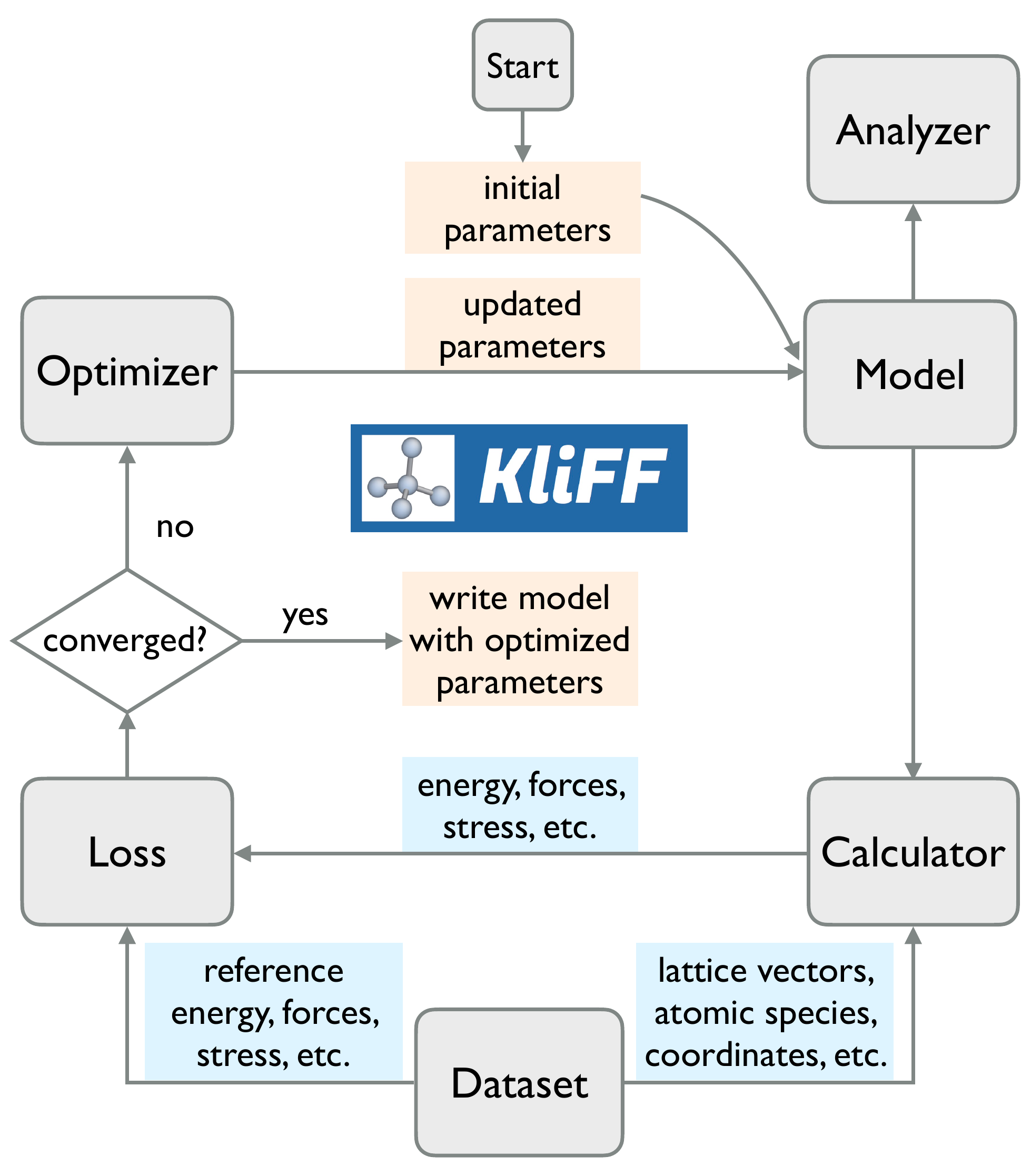}
\caption{Flowchart of the procedures of using KLIFF to train an IP.}
\label{fig:flowchart}
\end{figure}

KLIFF is written primarily in Python with several computationally intensive
components implemented in C++ accessible via Python bindings.
As such, users interact with KLIFF through a pure Python interface.
KLIFF is built in a modular fashion, as discussed in \sref{sec:uniform},
with key modules \verb|Dataset|, \verb|Model|,
\verb|Calculator|, \verb|Loss|, \verb|Optimizer|, and \verb|Analyzer|.
A flowchart showing the interaction and information transfer between
these modules for IP training is displayed in \fref{fig:flowchart}.
The modules are described below.

% flowchart

\subsection{Dataset module}
A dataset is comprised of a set of atomic configurations, which provide the
training data to optimize IP parameters or provide the test data to test the
quality of an IP\@.
An atomic configuration includes three vectors defining the simulation
cell, flags to indicate whether periodic boundary conditions (PBCs) are
applied along the cell vectors,
the species and coordinates of all atoms in the configuration,
and reference outputs.
KLIFF reads atomic configurations from extended XYZ files,
with each configuration stored in a separate file. The reference outputs
(energy, forces, and stress) associated with an atomic configuration are
also read in from the extended XYZ file. The standard XYZ format only stores
the number of atoms in a configuration and the species and
coordinates of the atoms. The extended XYZ format allows for additional information
to be stored, either in the second line via a series of \verb|key=value| pairs
(e.g.\ \verb|PBC="T,T,T"| and \verb|energy=1.2|) or in the body section by appending values
(e.g.\ forces) to the coordinates. Internally, each atomic
configuration and the reference outputs are associated with a \verb|Configuration|
object and a \verb|Dataset| is essentially a set of Configuration objects.

\subsection{Model module}
The fitting process begins with the instantiation of a model (IP).
Depending on the nature of the model, different operations can be applied.
For KIM models and physics-based KLIFF potentials, KLIFF can provide information on what parameters are
available for fitting, together with a description of each parameter and the
data structure and data type of each parameter.
Based on this information, a user can select
the parameters to fit and specify initial values or use defaults.
Lower and upper bounds on parameter values can also be provided to restrict it
to a range.
For an NN model, the descriptor representation of an atomic environment,
which serves as the input to the NN model, must be defined.
Then the NN can be constructed using an arbitrary number of layers,
nodes per layer, and activation functions.
Unlike physics-based models, KLIFF automatically initializes the parameters in
the network.
For example, the He initializer \cite{He_2015} is used to initialize the weights
and biases in \eref{eq:nn:connect}.
Other default choices are made by KLIFF based on the authors' physical understanding
and experience
to make it easier for users to develop machine learning potentials without
having to master subtle aspects of machine learning training.
For example, in a standard dropout approach, different NN connections would be removed
for each atom in a configuration (see \sref{sec:uncertainty}). However, KLIFF defaults to
a native dropout scheme that removes the same NN connections for all atoms in a configuration.
This ensures that atoms with identical environments (e.g.\ all atoms in an ideal silicon crystal)
will have the same atomic energy, forces, and other properties.
Users can overwrite default choices, for example, by selecting the native PyTorch
dropout instead of KLIFF's native implementation.

\subsection{Calculator module}
The created model is attached to a \emph{calculator}
that computes the predictions corresponding to the
reference outputs for the atomic configurations in the training set.
The native KLIFF calculator supports the evaluation of energy, forces,
and stress. If a property other than these is to be fitted, a new calculator
needs to be implemented. A new calculator can wrap any KIM compliant molecular
simulation package to compute the property with the given model in a similar
fashion to ASE calculators \cite{larsen2017atomic,ase}.

\subsection{Loss module}
The predictions computed by the calculator and the corresponding reference
output values stored in the training set are then used to construct a loss
function (e.g.\ \eref{eq:loss}) that quantifies the difference between the model
predictions and the references.
A weight can be assigned to each configuration, so that ``important''
configurations are emphasized more during optimization.
If the available loss functions in KLIFF do not satisfy a specific need, a user-defined loss
function can be added.

\subsection{Optimizer module}
The optimization process involves minimization of the loss function
with respect to the IP parameters until specified stopping criteria
are satisfied, such as reducing the loss function value below a tolerance
or reaching a maximum allowed number of minimization steps.

The optimizers supported by KLIFF can be broadly categorized in two classes:
\emph{batch optimizers} and \emph{mini-batch optimizers}.
The former (e.g.\ the L-BFGS-B and geodesic LM methods) typically require the evaluation
of the entire training set at each minimization step, whereas the latter
(e.g.\ the SGD and Adam methods) only use a subset of the
training set at a time.
Batch optimizers guarantee a monotonic decrease of the loss throughout the
minimization process and typically yield smaller
final loss values compared with mini-batch optimizers.
Mini-batch optimizers become advantageous for very large training sets
(typical of machine learning potentials) where evaluation of the entire training set
becomes prohibitive due to memory and/or computing constraints.
For NN models that contain a large number of parameters, SGD-based
optimizers can typically find a reasonable solution in parameter space that minimizes
the loss to a certain level.
By default,
KLIFF uses an L-BFGS-B optimizer for physics-based potentials, which typically have relatively
small numbers of parameters and small training sets, and an SGD-based Adam optimizer
for NN potentials, which have many parameters and very large training sets.
The user can overwrite this default and select a preferred optimizer.

Once the optimization is completed,
the fitted IP can be written out as a KIM model that conforms to the
KIM API, which can then be run against KIM VCs and KIM tests or be
used with any KIM-compliant simulations codes as discussed in \sref{sec:integ:kim}.
Generated KIM models, can be uploaded to \url{https://openkim.org} to receive a DOI
and make the model available to the broader research community.
Also, the model can be attached to an \verb|Analyzer| to carry out post-processing
analysis, such as computing the FIM discussed in \sref{sec:uncertainty}
and computing the root mean square errors of energy and forces for a test set.

\subsection{Command line tool}
KLIFF provides a command line tool called \verb|kliff| that facilitates
the execution of many common tasks.
For example, query a physics-based potential for available parameters
that can be optimized and their associated metadata,
print a synopsis of the atomic configurations in the dataset, or split
a dataset into multiple subsets.
Once installed, executing ``\verb|kliff --help|'' in the terminal
will list the commands, their arguments, and help information.

%%%%%%%%%%%%%%%%%%%%%%%%%%%%%%%%%%%%%%%%%%%%%%%%%%%%%%%%%%%%%%%%%%%%%%%%%%%%%%%%
% demo
%%%%%%%%%%%%%%%%%%%%%%%%%%%%%%%%%%%%%%%%%%%%%%%%%%%%%%%%%%%%%%%%%%%%%%%%%%%%%%%%
\section{Demonstration}
\label{sec:demo}

KLIFF has been extensively tested through the development of multiple IPs,
including an SW potential for two-dimensional molybdenum disulfide \cite{wen2017sw}, an interlayer potential
for multilayer graphene \cite{wen2018dihedral}, a hybrid NN potential for multilayer graphene \cite{wen2019hybrid},
and a dropout uncertainty NN potential (DUNN) to quantify uncertainty in molecular simulations \cite{wen2020uncertainty}.
In this section we present examples demonstrating the use of KLIFF
in training an SW potential and an NN potential for silicon.
The functional forms of the two IPs are described in \sref{sec:ip}.

%%%%%%%%%%%%%%%%%%%%%%%%%%%%%%%%%%%%%%%%
% training
%%%%%%%%%%%%%%%%%%%%%%%%%%%%%%%%%%%%%%%%
\subsection{Parameterization}
\label{sec:parameterization}

The training set is comprised of the energies and forces
for 2513 configurations of silicon in the diamond cubic crystal structure.
This includes configurations with compressed and stretched cells and
random perturbations of atoms, as well as configurations drawn from
a molecular dynamics trajectory at a temperature of $300~\text{K}$.
Since this is only a demonstration, instead of using first-principles
calculation or experimental data, the configurations were generated using
the EDIP model \cite{MD_506186535567_002, MD_506186535567_002a, MD_506186535567_002b, MD_506186535567_002c}.
%The dataset is randomly divided into a training set of 2262 configurations
%(90\%) and a test set of 251 configurations (10\%).
The dataset is provided in the supplementary material \cite{supplementary}.

The SW potential has seven parameters,
$A, B, p, q, \sigma, \lambda, \gamma$, along with the cutoff radius
$r^\text{cut}$ and the equilibrium angle $\beta^0$.
The cutoff radius is set to $r^\text{cut} = 3.77118~\text{\AA}$,
as used by Stillinger and Weber \cite{stillinger1985computer}, and the
equilibrium angle is set to the tetrahedral angle of the ideal cubic diamond structure,
 $\beta^0 = 109.47^\circ$.
Following most SW parameterizations
\cite{stillinger1985computer,wen2017sw,zhou2013stillinger}, the parameters $p$
and $q$ are set to 4 and 0, respectively.
The values of the remaining parameters are obtained by minimizing the loss function
in \eref{eq:loss} using the geodesic LM algorithm
\cite{transtrum2011geometry,transtrum2012geodesic,transtrum2012improvements}.
The energy and force weights are set to $w_m^\text{e} = 1/(N_m)^2$
and $w_m^\text{f} = 10/(N_m)^2$.
A larger force weight is used to better reproduce the phonon dispersions discussed in \sref{sec:test:ip}.
One exception is that the energy weight is set to $w_m^\text{e} = 10/(N_m)^2$
for configurations that have an ideal cubic diamond structure at different
lattice parameters. The increased weight ensures that these configurations
are not underrepresented in the fitting since
their force terms in \eref{eq:loss}
are identically zero (regardless of the IP parameters) due to
the symmetry of the underlying structure.
The optimal parameter set identified by this process and the preset parameters
are listed in \tref{tab:sw:params}.

% sw params
\begin{table}
\centering
\caption{Summary of SW parameters obtained by minimizing the loss function and the preset parameters.}
\label{tab:sw:params}
\begin{tabular}{cccc}
\hline
\hline
Parameter        &Value    &Parameter              &Value  \\
\hline
 $A$             &15.46588611~eV               &$B$               &0.61032816 \\
 $p$             &4                            &$q$               &0  \\
 $\sigma$        & 2.05971554~\AA              &$\lambda$         & 65.46736831~eV  \\
 $\gamma$        & 2.71009995~\AA              &$r^\text{cut}$    &3.77118~\AA  \\
 $\beta^0$ &$109.47^\circ$  & & \\
\hline
\end{tabular}
\end{table}

For the NN potential, we employ the $G_i^2$ and $G_i^5$ symmetry functions
(Eqs.~\eqref{eq:g2} and \eqref{eq:g5}) as the descriptors for characterizing
atomic environments.
The hyperparameters $\alpha$ and $R_\text{s}$ in \eref{eq:g2} and $\zeta, \lambda$, and $\eta$ in \eref{eq:g5} are provided
in the supplementary material \cite{supplementary}.
The cutoff in \eref{eq:sym_fn:cutoff} is set to $r^\text{cut} =
3.5~\text{\AA}$ to include only nearest-neighbor interactions.
A challenging aspect of training an NN, which is also a source of the
power and flexibility of the method, is that it is up to the developer to
select the number of descriptor terms to retain, the number of hidden
layers, the number of nodes within each hidden layer (which need
not be the same), and the activation function.
It is also possible to create different connectivity scenarios between layers.
Here we have opted for simplicity and adopted a fully-connected network with the
same number of nodes in each hidden layer.
The number of hidden layers and the number of nodes in each hidden layers are
determined through a grid search and are listed in \tref{tab:nn:params}.
The activation function $\act$ is taken to be the commonly used
hyperbolic tangent function, $\tanh(x) = (e^x - e^{-x})/(e^x + e^{-x})$.

The NN potential parameters
are obtained by minimizing the loss function \eref{eq:loss}.
The energy weight $w_m^\text{e}$ and forces weight $w_m^\text{f}$ are the
same as those used for the SW potential.
The minimization is carried out using the Adam optimizer
\cite{kingma2014adam} with a learning rate of 0.001.
As discussed in \sref{sec:implementation}, to accelerate the training process
a mini-batch technique \cite{li2014efficient} is employed with a
batch size of 100 configurations at each minimization step for a total of
2000 epochs.\footnote{An epoch is one complete pass over the dataset.
For example, if a dataset includes 50 configurations and a
mini-batch size of 10 configuration is used,
then one epoch consists of 5 minimization steps.}
%See \tref{tab:nn:params} for a summary of the NN parameters.

% NN param
\begin{table}
\centering
\caption{Summary of parameters in the NN potential and hyperparameters that
define the NN structure.}
\label{tab:nn:params}
\begin{tabular}{cc}
\hline
\hline
number of hidden layers              &3  \\
number of nodes in hidden layers     &10 \\
cutoff $r^\text{cut}$                &3.5~\AA \\
activation function $\act$           &$\tanh$ \\
descriptor hyperparameters           &see \cite{supplementary} \\
%weights                              &see SM \cite{supplementary} \\
%biases                               &see SM \cite{supplementary} \\
\hline
\end{tabular}
\end{table}

The scripts used to train the SW and NN potentials are provided in the
supplementary material \cite{supplementary}.

%%%%%%%%%%%%%%%%%%%%%%%%%%%%%%%%%%%%%%%%
% testing
%%%%%%%%%%%%%%%%%%%%%%%%%%%%%%%%%%%%%%%%
\subsection{Testing the trained potentials}
\label{sec:test:ip}

To test the fitted SW and NN potentials, we applied them to study energetic
and vibrational properties of silicon in the diamond cubic crystal structure.
As discussed in \sref{sec:integ:kim}, IPs trained by KLIFF can be exported
in a form compatible with the KIM API, which allows them to be used directly
with a variety of major molecular simulation packages,
such as LAMMPS \cite{plimpton1995fast,lammps2021,lammps}.
The tests described in this section were carried out using LAMMPS.

% energy vs. alat
\begin{figure}
\centering
\includegraphics[width=0.9\columnwidth]{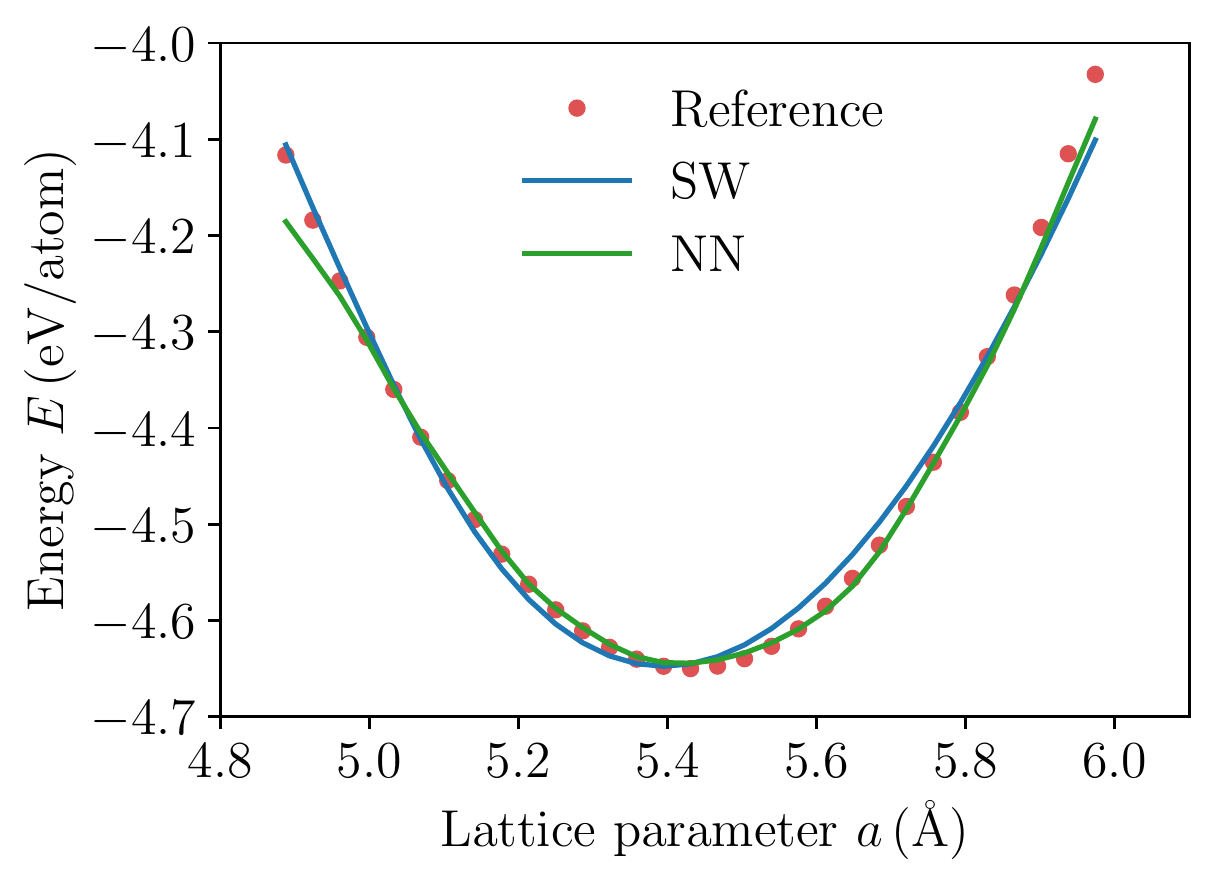}
\caption{Total potential energy of silicon as a function of the lattice parameter
predicted by the trained SW and NN potentials along with the EDIP reference data.}
\label{fig:e:vs:a}
\end{figure}

\begin{table}
\centering
\caption{Cohesive energy (absolute value of the minimum of the energy versus lattice
parameter curve) and equilibrium lattice constant for the diamond structure computed
using the EDIP potential (taken as the reference) and the SW and NN potentials (with
errors relative to EDIP given in parentheses).}
\begin{tabular}{lll}
\hline
\hline
Potential & $E_{\rm coh}$ [eV/atom] & $a_0$ [\AA] \\
\hline
EDIP & 4.650 & 5.43 \\
SW & 4.647 (0.06\%) & 5.39 (0.74\%) \\
NN & 4.645 (0.1\%) & 5.42 (0.18\%) \\
\hline
\end{tabular}
\label{tab:equil}
\end{table}

First, we investigate the cohesive energy versus lattice parameter for
ideal cubic diamond silicon (see structure in \fref{fig:diamond:si}).
The fitted SW and NN potentials are compared with the EDIP reference
data in \fref{fig:e:vs:a}. Both potentials reproduce the equilibrium
state well as seen in \tref{tab:equil}, however the NN potential with its
flexible functional form is able to follow the reference data more closely
across most of the range except for lattice parameters smaller than 5~\AA\
and larger than 5.9~\AA\@.
The training set contains configurations with lattice parameters up to
$\pm 10\%$ from the equilibrium value (i.e.\ $4.89 \sim 5.97~\text{\AA}$).
Thus configurations with lattice parameters smaller than 5~\AA\ and larger
than 5.9~\AA\ are at the ``edge'' of the training data where accuracy of the
NN potential is clearly reduced.
This is consistent with the discussion in \sref{sec:intro}. While highly
accurate within the training set, the NN potential has low transferability and
thus its ability to extrapolate beyond its training set is limited. This is
particularly clear on the compressive end of the response (lattice constant
smaller than 5.0~\AA). In contrast, the SW potential has a lower accuracy
overall since it is constrained by its physical functional form, but this
leads to a more correct trend outside the training set.

% dunn energy vs alat
\begin{figure}[t]
\centering
\includegraphics[width=0.9\columnwidth]{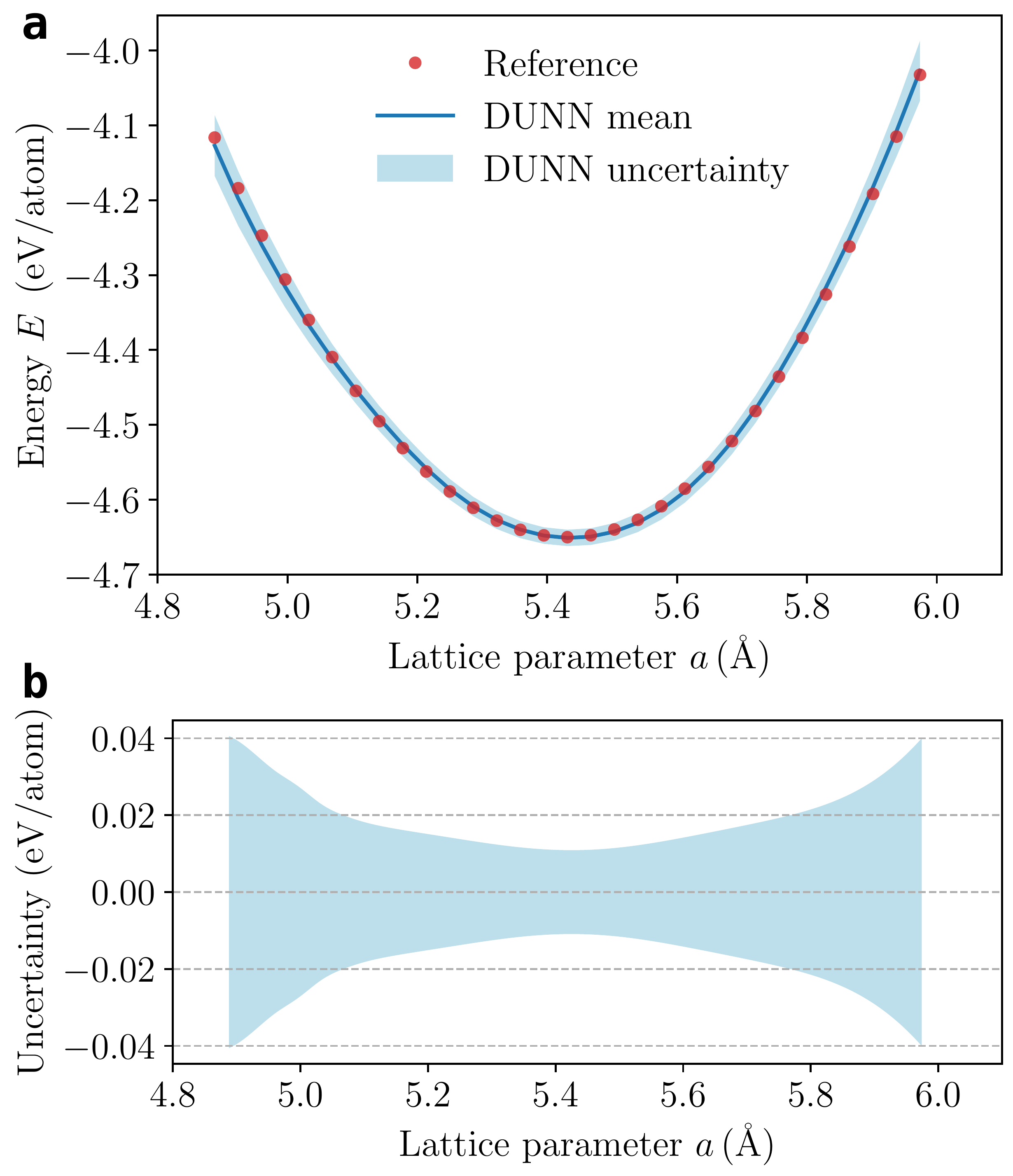}
\caption{Energy of silicon as a function of the lattice parameter predicted by the DUNN potential.
(a) Predictive mean and uncertainty of the energy by DUNN, where the uncertainty band is  twice the width of the standard deviation in the energy. Also plotted are the reference EDIP energies.
(b) DUNN uncertainty. The uncertainty band is the same as that in panel (a) except that here it is centered around 0 instead of the prediction mean in panel (a).
}
\label{fig:dunn:e:vs:a}
\end{figure}

It is important to quantify the uncertainty in the predictions of machine learning
potentials given their low transferability. As discussed in \sref{sec:uncertainty},
KLIFF supports the training of DUNN potentials \cite{wen2020uncertainty} that are
based on dropout uncertainty estimation. To demonstrate this, we train a DUNN potential
for the silicon dataset and apply it to investigate the same energy versus lattice
parameter problem discussed above. Since the emphasis is on the uncertainty in
energy, forces are not used in the training. (Details of the parameterization procedure
are provided in the supplementary material \cite{supplementary}.)
When a DUNN model is used it provides a mean value, which is the average over
the dropout ensemble, and an associated uncertainty estimate. The results for the
cohesive energy versus lattice parameter are compared with the EDIP reference data
in \fref{fig:dunn:e:vs:a}(a)). The mean DUNN values are in excellent agreement
with the reference data.\footnote{The agreement is better than the NN potential
in \fref{fig:diamond:si} since only energies are used in training the DUNN potential
allowing it to obtain a better fit, whereas the NN potential is fit using energies
and forces.} More importantly, the band around the mean values shows that the DUNN
uncertainty estimate increases as the silicon crystal is strained away from its
equilibrium state ($a=5.43~\text{\AA}$ and that the increase accelerates
towards the edges of the training set (see \fref{fig:dunn:e:vs:a}(b)).
Such uncertainty information can help to determine whether a molecular simulation
is reliable or not.

% phonon
\begin{figure}
\centering
\includegraphics[width=1.0\columnwidth]{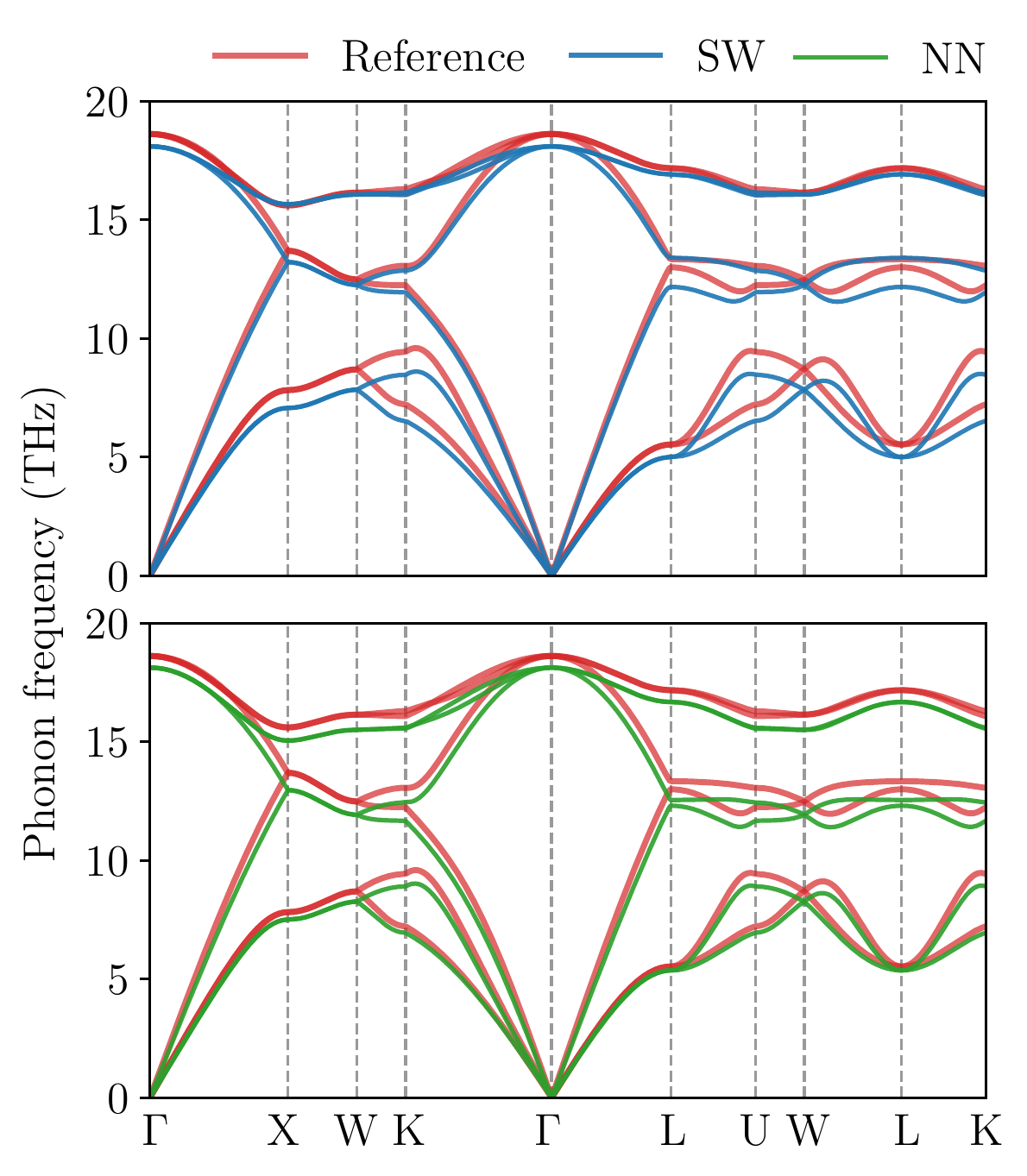}
\caption{Phonon dispersions of diamond cubic silicon along high symmetry points in the first Brillouin zone predicted by the trained SW and NN potentials along with the reference data by EDIP.}
\label{fig:phonon}
\end{figure}

As a second example, we consider phonon dispersion.
This set of curves provides a comprehensive view
of the elastic vibrational properties of a material, which play a key role in
many dynamical properties including thermal transport and stress wave
propagation.
It is therefore important for IPs to predict phonon dispersion correctly.
\fref{fig:phonon} presents the phonon dispersion curves of silicon
along high-symmetry points in the first Brillouin zone obtained using the
phonopy package \cite{phonopy}.
%The phonon frequency of all branches by the NN potential are a bit smaller than the
%reference EDIP prediction, and there appears to be a universal scale
%of 1.027 between them (i.e.\ if we multiply the NN prediction by 1.027, it will
%fall on top of the reference data).
The SW potential is in better agreement with the reference data for
branches with larger phonon frequencies,
but is less accurate for the two lowest-frequency branches, which can be
seen at the W, K, and U points.
Despite these small differences, the predictions by both the SW and NN potentials are in good agreement with the reference data\@.
The training set does not explicitly contain phonon frequency data, so the fact that both the SW
and NN potentials are able to correctly reproduce EDIP's phonon dispersion
curves indicates that they provide a good representation for the EDIP functional
form near the equilibrium state.

%%%%%%%%%%%%%%%%%%%%%%%%%%%%%%%%%%%%%%%%%%%%%%%%%%%%%%%%%%%%%%%%%%%%%%%%%%%%%%%%
\section{Summary and outlook}
\label{sec:summary}

In this paper, we introduce the \emph{KIM-based learning-integrated fitting
framework} (KLIFF) for developing IPs.
KLIFF provides a uniform Python user interface to train both physics-based and
machine learning potentials. It is flexible and easily extended to
support new atomic environment descriptors, models, loss functions, minimizers,
and analyzers.
KLIFF integrates closely with the KIM framework. An IP trained using KLIFF can be readily
deployed in a format consistent with the KIM API, which enables it to be used directly
in major simulation codes such as LAMMPS \cite{plimpton1995fast,lammps2021,lammps},
ASE \cite{larsen2017atomic,ase}, DL\_POLY \cite{smith1996,dlpoly}, GULP
\cite{gale1997gulp,gulp} and ASAP \cite{asap} among others.
The package is distributed under an open-source license and is available at
\url{https://github.com/openkim/kliff} along with a
comprehensive user manual with several tutorials.

KLIFF (version 0.3.0) is fully functional as demonstrated in this paper
by training the SW, NN, and DUNN potentials for silicon.
Development continues with an emphasis on incorporating new features,
including
(1) supporting more machine learning models and descriptors;
(2) integration with KIM tests to train on material properties beyond energy,
forces, and stress; and
(3) creation of tools for automatic selection of hyperparameters for machine
learning potentials (e.g.\ optimal number of terms to retain for a descriptor and
optimal number of layers and nodes in each layer for an NN potential).
We encourage other researchers to contribute to the development, and
provide full and detailed documentation of the KLIFF API
(see the Package Reference section in the documentation \url{https://github.com/openkim/kliff}).

%%%%%%%%%%%%%%%%%%%%%%%%%%%%%%%%%%%%%%%%%%%%%%%%%%%%%%%%%%%%%%%%%%%%%%%%%%%%%%%%
\section*{Acknowledgements}
This research was partly supported by the Army Research Office (W911NF-14-1-0247)
under the MURI program, the National Science Foundation (NSF) under grants
DMR-1834251, DMR-1834332 and OAC-2039575, and through the University of Minnesota MRSEC
under Award Number DMR-1420013.
The authors wish to acknowledge the Minnesota Supercomputing Institute (MSI) at the
University of Minnesota for providing resources that contributed to the results
reported in this paper.
MW thanks the University of Minnesota Doctoral Dissertation Fellowship for
supporting his research.



%%%%%%%%%%%%%%%%%%%%%%%%%%%%%%%%%%%%%%%%%%%%%%%%%%%%%%%%%%%%%%%%%%%%%%%%%%%%%%%%
%% References

\bibliographystyle{elsarticle-num}
%\bibliography{kliff.bib}






% \end{thebibliography}

\clearpage


\section*{Supplementary material}

# Supplementary material for "KLIFF: A framework to develop physics-based and machine learning interatomic potentials"

Mingjian Wen[a], Yaser Afshar[a], Ryan S. Elliott[a], Ellad B. Tadmor[a,*]

[a]*Department of Aerospace Engineering and Mechanics, University of Minnesota, Minneapolis, MN 55455, USA*

## 1. Parameterization of a DUNN potential

The architecture of the dropout uncertainty neural network (DUNN) potential is the same as the neural network (NN) potential except that dropout layers are added after the activation layers. We use a dropout ratio of 0.1 and 64 nodes in each layers. As briefly discussed in the main text, instead of both energies and forces, we only include energies for training the DUNN potential to focus on investigating the uncertainty in the energy predictions. Other parameterization choices (e.g. atomic environment descriptor and cutoff distance) are exactly the same as those for the NN potential discussed in the main text. An example training script is given in Section 2.3.

## 2. Dataset and code examples

We provide the scripts used for training the Stillinger-Weber (SW), NN, and DUNN potentials. These scripts are compatible with KLIFF v.0.3.0 and may require changes in future releases of KLIFF. For up-to-date examples, see the KLIFF documentation at https://github.com/openkim/kliff. The dataset used to fit the potentials is provided along with this supplementary material.

### 2.1. KLIFF script for training an SW potential

```python
from kliff.calculators import Calculator
from kliff.dataset import Dataset
from kliff.loss import Loss
from kliff.models import KIMModel

# Create a model for the SW potential and set parameters to optimize
model = KIMModel(model_name="SW_StillingerWeber_1985_Si__MO_405512056662_006")
model.set_opt_params(
    A=[["default"]], B=[["default"]], sigma=[["default"]], gamma=[["default"]]
)
model.set_one_opt_param("lambda", [["default"]])

# Load training set
dataset_name = "Si_dataset_EDIP"
tset = Dataset(dataset_name)
configs = tset.get_configs()

# Set weight for configurations without perturbation to 10 (default is 1)
for conf in configs:
    if "perturb0" in conf.identifier:
        conf.weight = 10
```

---

*Corresponding author.
*E-mail address:* tadmor@umn.edu

```python
# Create calculator for the SW model
calc = Calculator(model)
calc.create(configs)

# Minimize using the geodesic Levenberg-Marquardt optimizer on 4 processes
loss = Loss(calc, residual_data={"energy_weight": 1, "forces_weight": 10}, nprocs=4)
loss.minimize(method="geodesiclm", maxiter=1000)

# Save model for future reload and write a KIM model
model.save("kliff_model.yaml")
model.write_kim_model()
```

*2.2. KLIFF script for training an NN potential*

```python
from kliff import nn
from kliff.calculators import CalculatorTorch
from kliff.dataset import Dataset
from kliff.descriptors import SymmetryFunction
from kliff.loss import Loss
from kliff.models import NeuralNetwork

# Initialize the NN model using the symmetry functions descriptor
descriptor = SymmetryFunction(
    cut_name="cos", cut_dists={"Si-Si": 3.5}, hyperparams="set51", normalize=True
)
model = NeuralNetwork(descriptor)

# Create an NN model of 3 hidden layers with 10 nodes in each
N = 10
model.add_layers(
    # first hidden layer
    nn.Linear(descriptor.get_size(), N),
    nn.Tanh(),
    # second hidden layer
    nn.Linear(N, N),
    nn.Tanh(),
    # third hidden layer
    nn.Linear(N, N),
    nn.Tanh(),
    # output layer
    nn.Linear(N, 1),
)

# Load training set
dataset_name = "Si_dataset_EDIP"
tset = Dataset(dataset_name)
configs = tset.get_configs()

# Set weight for configurations without perturbation to 10 (default is 1)
for conf in configs:
    if "perturb0" in conf.identifier:
        conf.weight = 10

# Create calculator for the NN model
calc = CalculatorTorch(model)
calc.create(configs)

# Minimizing for a maximum of 2000 epochs
loss = Loss(calc, residual_data={"energy_weight": 1, "forces_weight": 10})
loss.minimize(method="Adam", num_epochs=2000, batch_size=100, lr=0.001)

# Save model for future reload and write a KIM model
model.save("kliff_model.pkl")
loss.save_optimizer_state("optimizer_stat.pkl")
```

```
model.write_kim_model()
```

*2.3. KLIFF script for training a DUNN potential*

```python
from kliff import nn
from kliff.calculators import CalculatorTorch
from kliff.dataset import Dataset
from kliff.descriptors import SymmetryFunction
from kliff.loss import Loss
from kliff.models import NeuralNetwork

# Initialize the NN model using the symmetry functions descriptor
descriptor = SymmetryFunction(
    cut_name="cos", cut_dists={"Si-Si": 3.5}, hyperparams="set51", normalize=True
)
model = NeuralNetwork(descriptor)

# Create an NN model of 3 hidden layers with 10 nodes in each
N = 64
model.add_layers(
    # first hidden layer
    nn.Linear(descriptor.get_size(), N),
    nn.Tanh(),
    nn.Dropout(p=0.1),
    # second hidden layer
    nn.Linear(N, N),
    nn.Tanh(),
    # nn.Dropout(p=0.1),
    # third hidden layer
    nn.Linear(N, N),
    nn.Tanh(),
    # nn.Dropout(p=0.1),
    # output layer
    nn.Linear(N, 1),
)

# Load training set
dataset_name = "../Si_dataset_EDIP_child-1"
tset = Dataset(dataset_name)
configs = tset.get_configs()

# Set weight for configurations without perturbation to 10 (default is 1)
for conf in configs:
    if "perturb0" in conf.identifier:
        conf.weight = 10

# Create calculator for the DUNN model and turn off the calculation of forces
calc = CalculatorTorch(model)
calc.create(configs, use_forces=False)

# Minimizing for a maximum of 2000 epochs
loss = Loss(calc, residual_data={"energy_weight": 1, "forces_weight": 0})
loss.minimize(method="Adam", num_epochs=2000, batch_size=100, lr=0.001)

# Save model for future reload and write a KIM model
model.save("kliff_model.pkl")
loss.save_optimizer_state("optimizer_stat.pkl")
model.write_kim_model(dropout_ensemble_size=100)
```

## 3. Symmetry function hyperparameters

The atomic environment descriptor for the NN and DUNN potentials is obtained using the symmetry functions $G_i^2$ and $G_i^4$. The hyperparameters used in $G_i^2$ are listed in Table 1 and in $G_i^4$ are listed in Table 2.

Table 1: Hyperparameters used in the radical descriptor $G_i^2$.

| No. | $\alpha$ (Bohr$^{-2}$) | $R_s$ (Bohr) |
|---|---|---|
| 1 | 0.001 | 0 |
| 2 | 0.01 | 0 |
| 3 | 0.02 | 0 |
| 4 | 0.035 | 0 |
| 5 | 0.06 | 0 |
| 6 | 0.1 | 0 |
| 7 | 0.2 | 0 |
| 8 | 0.4 | 0 |

Table 2: Hyperparameters used in the angular descriptor $G_i^4$.

| No. | $\zeta$ | $\lambda$ | $\eta$ (Bohr$^{-2}$) | No. | $\zeta$ | $\lambda$ | $\eta$ (Bohr$^{-2}$) |
|---|---|---|---|---|---|---|---|
| 1 | 1 | −1 | 0.0001 | 23 | 2 | −1 | 0.025 |
| 2 | 1 | 1 | 0.0001 | 24 | 2 | 1 | 0.025 |
| 3 | 2 | −1 | 0.0001 | 25 | 4 | −1 | 0.025 |
| 4 | 2 | 1 | 0.0001 | 26 | 4 | 1 | 0.025 |
| 5 | 1 | −1 | 0.003 | 27 | 16 | −1 | 0.025 |
| 6 | 1 | 1 | 0.003 | 28 | 16 | 1 | 0.025 |
| 7 | 2 | −1 | 0.003 | 29 | 1 | −1 | 0.045 |
| 8 | 2 | 1 | 0.003 | 30 | 1 | 1 | 0.045 |
| 9 | 1 | −1 | 0.008 | 31 | 2 | −1 | 0.045 |
| 10 | 1 | 1 | 0.008 | 32 | 2 | 1 | 0.045 |
| 11 | 2 | −1 | 0.008 | 33 | 4 | −1 | 0.045 |
| 12 | 2 | 1 | 0.008 | 34 | 4 | 1 | 0.045 |
| 13 | 1 | −1 | 0.015 | 35 | 16 | −1 | 0.045 |
| 14 | 1 | 1 | 0.015 | 36 | 16 | 1 | 0.045 |
| 15 | 2 | −1 | 0.015 | 37 | 1 | −1 | 0.08 |
| 16 | 2 | 1 | 0.015 | 38 | 1 | 1 | 0.08 |
| 17 | 4 | −1 | 0.015 | 39 | 2 | −1 | 0.08 |
| 18 | 4 | 1 | 0.015 | 40 | 2 | 1 | 0.08 |
| 19 | 16 | −1 | 0.015 | 41 | 4 | −1 | 0.08 |
| 20 | 16 | 1 | 0.015 | 42 | 4 | 1 | 0.08 |
| 21 | 1 | −1 | 0.025 | 43 | 16 | 1 | 0.08 |
| 22 | 1 | 1 | 0.025 | | | | |

## 4. Implementation of a new potential in KLIFF

KLIFF is designed to be extendable. This example demonstrates how to add a new interatomic potential of the Lennard-Jones form.

```python
from typing import Dict

import numpy as np
from kliff.dataset.configuration import Configuration
from kliff.models.model import ComputeArguments, Model
from kliff.models.parameter import Parameter
from kliff.neighbor import NeighborList, assemble_forces


class LennardJones(Model):
    """
    Lennard-Jones 6-12 potential model.
    """

    def __init__(self, model_name: str = "LJ6-12"):
        super(LennardJones, self).__init__(model_name)

    def init_model_params(self):
        model_params = {
            "epsilon": Parameter([15.3]),
            "sigma": Parameter([0.01025]),
            "cutoff": Parameter([3.5]),
        }

        return model_params

    def init_influence_distance(self):
        return self.model_params["cutoff"][0]

    def init_supported_species(self):
        return {"Ar": 1}

    def get_compute_argument_class(self):
        return LJComputeArguments


class LJComputeArguments(ComputeArguments):
    """
    Lennard-Jones 6-12 potential computation routine.
    """

    implemented_property = ["energy", "forces"]

    def __init__(
        self,
        conf: Configuration,
        supported_species: Dict[str, int],
        influence_distance: float,
    ):
        super(LJComputeArguments, self).__init__(
            conf, supported_species, influence_distance
        )

        self.neigh = NeighborList(
            self.conf, influence_distance, padding_need_neigh=False
        )

    def compute(self, params: Dict[str, Parameter]):
        epsilon = params["epsilon"][0]
        sigma = params["sigma"][0]
        rcut = params["cutoff"][0]
        coords = self.conf.coords

        coords_including_padding = self.neigh.coords
```

5

```python
            forces_including_padding = np.zeros_like(coords_including_padding)

        energy = 0
        for i, xyz_i in enumerate(coords):
            neighlist, _, _ = self.neigh.get_neigh(i)
            for j in neighlist:
                xyz_j = coords_including_padding[j]
                rij = xyz_j - xyz_i
                r = np.linalg.norm(rij)
                phi, dphi = self.calc_phi_dphi(epsilon, sigma, r, rcut)
                energy += 0.5 * phi
                pair = 0.5 * dphi / r * rij
                forces_including_padding[i] = forces_including_padding[i] + pair
                forces_including_padding[j] = forces_including_padding[j] - pair

        self.results["energy"] = energy
        forces = assemble_forces(
            forces_including_padding, len(coords), self.neigh.padding_image
        )
        self.results["forces"] = forces

    @staticmethod
    def calc_phi_dphi(epsilon, sigma, r, rcut):
        if r > rcut:
            phi = 0.0
            dphi = 0.0
        else:
            sor = sigma / r
            sor6 = sor * sor * sor
            sor6 = sor6 * sor6
            sor12 = sor6 * sor6
            phi = 4 * epsilon * (sor12 - sor6)
            dphi = 24 * epsilon * (-2 * sor12 + sor6) / r
        return phi, dphi
```



\end{document}